\documentclass[12pt]{iopart}
\usepackage{graphicx}
\newcommand{\bvec}[1]{\mbox{\boldmath $#1$}}
\begin{document}

\title[Dimensional reduction in driven disordered systems]{Dimensional reduction in driven disordered systems}

\author{Taiki Haga}

\address{Department of Physics, Kyoto University, Kyoto 606-8502, Japan}
\ead{haga@scphys.kyoto-u.ac.jp}
\vspace{10pt}
\begin{indented}
\item[]\today
\end{indented}

\begin{abstract}
We investigate the critical behavior of disordered systems transversely driven at a uniform and steady velocity. 
An intuitive argument predicts that the long-distance physics of $D$-dimensional driven disordered systems at zero temperature is the same as that of the corresponding $(D-1)$-dimensional pure systems in thermal equilibrium.
This result is analogous to the well-known dimensional reduction property in thermal equilibrium, which states the equivalence between $D$-dimensional disordered systems and $(D-2)$-dimensional pure systems.
To clarify the condition that the dimensional reduction holds, we perform the functional renormalization group analysis of elastic manifolds transversely driven in random media.
We argue that the nonanalytic behavior in the second cumulant of the renormalized disorder leads to the breakdown of the dimensional reduction.
We further found that the roughness exponent is equal to the dimensional reduction value for the single component case, but it is not for the multi-component cases.
\end{abstract}

%

\section{Introduction}

It has been a long-standing problem in statistical mechanics to understand how the large-scale behaviors of pure systems are modified by quenched disorder.
One of the most remarkable achievements in this subject is the discovery of the so-called dimensional reduction property, which predicts that the long-distance physics of some disordered systems in spatial dimension $D$ is the same as that of the corresponding systems {\it without disorder} in a lower spatial dimension $D-2$ \cite{Aharony-76,Young-77,Grinstein-76,Parisi-79}.
Examples of such systems include elastic manifolds in random media \cite{Blatter-94,Nattermann-00,Gruner-98}, the random field Ising model (RFIM) \cite{Nattermann-97}, the random field and random anisotropy $O(N)$ models.
However, it is known that the dimensional reduction can break down due to a nonperturbative effect associated with the presence of multiple local minima in the energy landscape.
For example, although it predicts that the lower critical dimension of the RFIM is three, this contradicts the fact that the three-dimensional RFIM exhibits long-range order at weak disorder \cite{Imry-75,Imbrie-84,Bricmont-87}.
The breakdown of the dimensional reduction suggests that in disordered systems there is nontrivial physics beyond the standard perturbative field theory.

In addition to the equilibrium statistical mechanics of disordered systems, the structure and dynamics of disordered systems driven out of equilibrium have attracted considerable attention not only from theoretical interest but also from practical motivations.
This broad subject can be divided into two subtopics.
The first research topic is collective transports of many-body systems in random media \cite{Fisher-98,Fisher-85-1}, where a quantity of interest is the current-force characteristics, which describe how much current is induced by the external driving force.
Especially, most studies have been devoted to the dynamical critical phenomena in the quasi-static limit, such as depinning transitions and avalanche dynamics.
Renormalization group (RG) analysis of an elastic interface driven in a random potential elucidated the universal nature of these quasi-static behaviors \cite{Narayan-92,Narayan-93,Chauve-00,LeDoussal-02,LeDoussal-13}.

The second research topic is dynamical phase transitions in driven disordered systems \cite{Reichhardt-17}.
We ask what happens when interacting many-particle systems are driven in a random substrate.
The competition between the quenched disorder and driving force can lead to a novel type of phase transitions and critical phenomena which cannot be observed in thermal equilibrium.
For example, it is known that vortex lattices in dirty superconductors exhibit nonequilibrium phase transitions between various dynamical phases, e.g., liquid, smectic, moving glass, and so on, when they are driven by the Lorentz force \cite{Koshelev-94,Moon-96,Ryu-96,Dominguez-97,Yaron-94,Pardo-98}.
From theoretical viewpoints, there is an intrinsic difference between such nonequilibrium phase transitions and the depinning transitions mentioned above.
The latter are quasi-static phenomena where the criticality is reached in the zero limit of the driving velocity $v \to 0^+$, thus any dissipation does not take place.
In contrast, the former are truly nonequilibrium phenomena where the system is driven at a nonzero velocity and a steady state with a constant dissipation rate is realized.

Although there are a large number of theoretical studies for the quasi-static phenomena, little is understood about the long-distance physics of nonequilibrium steady states in driven disordered systems.
To provide a unified description of these systems, we are required to construct a field theoretical formalism including the effects of quenched disorder and nonequilibrium driving.
We expect that, if the standard perturbative treatment is applied to such a field theory, one might be led to incorrect results, as the dimensional reduction in the equilibrium statistical mechanics.
However, the perturbative prediction may provide a natural starting point to investigate new nonperturbative effects arising in nonequilibrium situations.
Therefore, we ask whether there is a dimensional reduction property for driven disordered systems, and if there is, how it breaks down.

In this study, we derive a dimensional reduction property for some class of  systems, which predicts that the critical behavior of $D$-dimensional driven disordered systems at zero temperature is the same as that of the corresponding $(D-1)$-dimensional pure systems in equilibrium.
However, this property does not always hold.
To demonstrate how it breaks down, we next investigate the large-scale behavior of an $N$-component elastic manifold transversely driven in a random potential by using the functional renormalization group (FRG) theory combined with a nonperturbative formalism.
We argue that the breakdown of the dimensional reduction is closely related to the nonanalytic behavior of the second cumulant of the renormalized random force.
The roughness exponent is calculated at $D=3-\epsilon$, and we find that it is equal to the dimensional reduction value for $N=1$ at all order of $\epsilon$, but it is not for the multi-component cases $N>1$.
Especially, we show that, for the short-range correlated disorder, the roughness exponent behaves as $(N+2)^{-1}$ for large $N$.

This paper is organized as follows.
In Sec.~\ref{sec:Model}, we introduce models, which describe elastic systems transversely driven at a uniform and steady velocity in random media.
We also remark the difference between our models and those discussed in the previous studies.
In Sec.~\ref{sec:Dimensional reduction}, we show that an intuitive argument predicts that the long-distance physics of these models is the same as that of the lower dimensional pure models.
We call this property the dimensional reduction.
However, this is not always correct.
Thus, we next perform the FRG analysis of driven random manifold model to elucidate the nonperturbative effect responsible for the breakdown of the dimensional reduction.
In Sec.~\ref{sec:NP-FRG formalism}, we sketch the nonperturbative implementation of the FRG theory, and the flow equation for the second cumulant of the renormalized disorder is derived.
In Sec.~\ref{sec:Results}, we numerically solve the flow equation.
The roughness exponent is calculated and its asymptotic behavior in the large $N$ limit is also discussed.
In Sec.~\ref{sec:Conclusions}, we provide concluding remarks.

\section{Model}
\label{sec:Model}

Let $\phi(\bvec{r})$ be a scalar or vector field characterizing the state of the system, such as the local magnetization in the Ising model or the position of an elastic interface.
We consider the following Hamiltonian in spatial dimension $D$:
\begin{eqnarray}
\mathcal{H}[\phi] = \int d^D \bvec{r} \left[ \frac{1}{2} |\nabla \phi(\bvec{r})|^2 + U(\phi(\bvec{r})) + V(\bvec{r};\phi(\bvec{r})) \right],
\label{General_Hamiltonian}
\end{eqnarray}
where $U(\phi)$ is a nonrandom potential and $V(\bvec{r};\phi)$ is a random potential.
The random force is then defined by 
\begin{equation}
F(\bvec{r};\phi)=-\partial_{\phi}V(\bvec{r};\phi).
\end{equation}
We assume that the random force obeys a mean-zero Gaussian distribution and its second cumulant is written as 
\begin{equation}
\overline{F(\bvec{r};\phi)F(\bvec{r}';\phi')}=\Delta(\phi-\phi')\delta(\bvec{r}-\bvec{r}'),
\label{bare_random_force_cumulant}
\end{equation}
where the over-bar represents the average over the disorder.
For the random manifold (RM), one assumes that $U(\phi)=0$, and for the RFIM, $U(\phi)=\lambda(\phi^2-\phi_0^2)^2$ and $\Delta(\phi)$ is set to a constant.

We next define the dynamics of the model.
The relaxation dynamics in thermal equilibrium is described by
\begin{eqnarray}
\partial_t \phi = -\frac{\delta \mathcal{H}[\phi]}{\delta \phi} + \xi.
\label{General_EM_eq}
\end{eqnarray}
The thermal noise $\xi$ satisfies 
\begin{equation}
\langle \xi(\bvec{r},t) \xi(\bvec{r}',t') \rangle =2 T \delta(\bvec{r}-\bvec{r'}) \delta(t-t'), 
\end{equation}
where $T$ is a temperature.
The stationary probability distribution of Eq.~(\ref{General_EM_eq}) is given by the Boltzmann distribution, $e^{-\mathcal{H}[\phi]/T}$.
We now add to Eq.~(\ref{General_EM_eq}) a term $\Lambda[\phi]$, which drives the system out of equilibrium.
We assume that $\Lambda[\phi]$ cannot be expressed as the functional derivative of a potential and it does not break the symmetry of $\phi$, such as the $\mathrm{Z}_2$ or $O(N)$ symmetry.
The simplest choice is that $\Lambda[\phi]=(\bvec{v} \cdot \nabla) \phi$, where $\bvec{v}$ is a constant vector.
Therefore, we define the dynamics as
\begin{eqnarray}
\partial_t \phi + v \partial_x \phi = -\frac{\delta \mathcal{H}[\phi]}{\delta \phi} + \xi,
\label{General_EM}
\end{eqnarray}
which describes the relaxation dynamics of interacting systems driven in random media with a uniform and steady velocity $\bvec{v}=(v,0,0)$.
For example, one can consider the dynamics of the transverse displacement field of the vortex lattices moving in dirty superconductors \cite{Giamarchi-96,LeDoussal-98,Balents-98} or that of liquid crystals flowing in porous media \cite{Araki-12}.

We consider the stationary state of Eq.~(\ref{General_EM}).
Let $P_{\mathrm{ss}}[\phi;V]$ be the stationary probability distribution function for a given random potential $V(\bvec{r};\phi)$.
Note that it is no longer written in the form of the Boltzmann distribution.
For an arbitrary functional $A[\phi]$, its average over the stationary state is written as
\begin{equation}
\langle A[\phi] \rangle = \int \mathcal{D} \phi A[\phi] P_{\mathrm{ss}}[\phi;V].
\end{equation}
Furthermore, its disorder average is defined by
\begin{equation}
\overline{\langle A[\phi] \rangle} = \int \mathcal{D}V \int \mathcal{D} \phi A[\phi] P_{\mathrm{ss}}[\phi;V] P_{\mathrm{R}}[V],
\label{Disorder_average_A}
\end{equation}
where $P_{\mathrm{R}}[V]$ is the probability distribution function of the random potential.

We remark the difference between our models and those discussed in the context of the depinning transition \cite{Chauve-00}.
The latter models are given by
\begin{eqnarray}
\partial_t \phi = -\frac{\delta \mathcal{H}[\phi]}{\delta \phi} + f + \xi,
\label{General_EM_longitudinal_driving}
\end{eqnarray}
where $U(\phi)=0$ and $f$ is a constant driving force.
Note that the driving force explicitly breaks the symmetry of $\phi$,
and in the depinning regime the average value of $\phi$ increases linearly in time.
On the other hand, in our models, the driving term $v \partial_x \phi$ respects the symmetry.
To distinguish our models from those discussed in Ref.~\cite{Chauve-00}, we call two types of driving described  by Eqs.~(\ref{General_EM}) and (\ref{General_EM_longitudinal_driving}) as ``transverse driving'' and ``longitudinal driving'', respectively.

\section{Dimensional reduction}
\label{sec:Dimensional reduction}

In equilibrium, standard perturbation theory predicts that the critical exponents of $D$-dimensional random field spin models are the same as those of $(D-2)$-dimensional pure spin models \cite{Aharony-76,Young-77,Grinstein-76,Parisi-79}.
Before we consider its nonequilibrium counterpart, we briefly review this conventional dimensional reduction.
Let $\phi_{\mathrm{st}}(\bvec{r})$ be a stationary state of the Hamiltonian,
\begin{equation}
\frac{\delta \mathcal{H}[\phi]}{\delta \phi} \biggr|_{\phi=\phi_{\mathrm{st}}} = 0.
\label{Stationary_equation_eq}
\end{equation}
The crucial point is that there are a large number of stationary states satisfying Eq.~(\ref{Stationary_equation_eq}).
We denote the set of all stationary states as $\{ \phi_{\mathrm{st}}^{(\chi)}(\bvec{r}) \}_{\chi=1,...,\mathcal{N}}$, where $\mathcal{N}$ is the number of the stationary states.
Note that $\phi_{\mathrm{st}}^{(\chi)}(\bvec{r})$ depends on the realization of the random potential $V(\bvec{r})$.
The correlation function is then given by
\begin{equation}
C(\bvec{r}) = \int \mathcal{D}V P_{\mathrm{R}}[V] \sum_{\chi=1}^{\mathcal{N}} \frac{e^{-\beta E_{\chi}}}{Z} \phi_{\mathrm{st}}^{(\chi)}(\bvec{r}) \phi_{\mathrm{st}}^{(\chi)}(\bvec{0}),
\label{correlation_eq}
\end{equation}
where $E_{\chi}=\mathcal{H}[\phi_{\mathrm{st}}^{(\chi)}]$ and $Z=\sum_{\chi=1}^{\mathcal{N}} e^{-\beta E_{\chi}}$.
More rigorously, one should distinguish minima, maxima, and saddle-points of the Hamiltonian, and the summation in Eq.~(\ref{correlation_eq}) should be taken over only local minima.
However, at low temperatures, since the contribution from the local minima is expected to be dominant, one does not need to care the restriction in the summation.
Especially, at zero temperature, $C(\bvec{r})$ can be expressed in terms of the ground state.
The calculation of Eq.~(\ref{correlation_eq}) is highly nontrivial.
Instead, we define a correlation function by averaging over $\{ \phi_{\mathrm{st}}^{(\chi)}(\bvec{r}) \}$ with the uniform weight,
\begin{equation}
C_{\mathrm{uni}}(\bvec{r}) = \int \mathcal{D}V P_{\mathrm{R}}[V] \frac{1}{\mathcal{N}} \sum_{\chi=1}^{\mathcal{N}} \mathrm{sign}(\chi) \phi_{\mathrm{st}}^{(\chi)}(\bvec{r}) \phi_{\mathrm{st}}^{(\chi)}(\bvec{0}),
\label{correlation_eq_uniform}
\end{equation}
where ``$\mathrm{sign}(\chi)$'' represents the sign of the Hessian $\mathrm{det}(\delta^2 \mathcal{H}/\delta \phi \delta \phi)$ evaluated at $\phi=\phi_{\mathrm{st}}^{(\chi)}$, for example, $\mathrm{sign}(\chi)=1$ for a local minimum of the Hamiltonian.
If there is only a single stationary state, $C(\bvec{r})=C_{\mathrm{uni}}(\bvec{r})$.
Remarkably, Eq.~(\ref{correlation_eq_uniform}) can be calculated by using the supersymmetry formalism and it is found to be identical to the correlation function of the pure system in $D-2$ dimensions  \cite{Parisi-79}.
However, the large-scale behavior of $C_{\mathrm{uni}}(\bvec{r})$ is not necessarily the same as that of the actual correlation function $C(\bvec{r})$.

Let us consider the driven disordered systems defined by Eq.~(\ref{General_EM}).
We show that an intuitive argument predicts that the critical behavior of the $D$-dimensional driven disordered systems at zero temperature is the same as that of the corresponding $(D-1)$-dimensional pure systems in equilibrium.
At zero temperature, the equation of motion is given by
\begin{equation}
\partial_t \phi(\bvec{r},t) + v \partial_x \phi(\bvec{r},t) = \nabla^2 \phi(\bvec{r},t) - U'(\phi(\bvec{r},t))+ F(\bvec{r};\phi(\bvec{r},t)).
\label{General_EM_zero_temp}
\end{equation}
The solution of Eq.~(\ref{General_EM_zero_temp}) reaches a stationary state $\phi_{\mathrm{st}}(\bvec{r})$ after a sufficiently long time.
We assume that there is only a single stationary state, which implies that $\phi_{\mathrm{st}}(\bvec{r})$ is independent of the initial condition of Eq.~(\ref{General_EM_zero_temp}).
The stationary state satisfies the following equation:
\begin{eqnarray}
v\partial_x \phi_{\mathrm{st}}(\bvec{r}) = \nabla^2 \phi_{\mathrm{st}}(\bvec{r}) - U'(\phi_{\mathrm{st}}(\bvec{r}))+ F(\bvec{r};\phi_{\mathrm{st}}(\bvec{r})).
\label{Stationary_equation}
\end{eqnarray}
In the large length scale, the longitudinal elastic term $\partial_x^2 \phi$ is negligible compared to the advection term $v\partial_x \phi$.
By ignoring the former term, we have the following equation:
\begin{eqnarray}
v\partial_x \phi(\bvec{r}) = \nabla_{\perp}^2 \phi(\bvec{r}) - U'(\phi(\bvec{r}))+ F(\bvec{r};\phi(\bvec{r})),
\label{Stationary_equation_perp}
\end{eqnarray}
where $\nabla_{\perp}$ is the derivative operator for the transverse directions.
One can easily obtain a solution of Eq.~(\ref{Stationary_equation_perp}) as follows.
First, we choose an arbitrary $(D-1)$-dimensional configuration $\phi(x=0,\bvec{r}_{\perp})$, where $\bvec{r}_{\perp}$ is the transverse coordinate.
Then, the whole solution $\phi(x,\bvec{r}_{\perp})$ can be obtained by propagating this ``initial'' configuration along $x$-direction according to Eq.~(\ref{Stationary_equation_perp}).
Although Eq.~(\ref{Stationary_equation_perp}) has infinitely many solutions, there exist a solution $\phi_{*}(x,\bvec{r}_{\perp})$ such that its large-scale behavior is the same as that of $\phi_{\mathrm{st}}(\bvec{r})$.
The important point is that $\phi_{*}(x,\bvec{r}_{\perp})$ is identical to a dynamical solution of the corresponding $(D-1)$-dimensional pure system {\it by considering the spatial coordinate $x$ as a fictitious time}.
Actually, Eq.~(\ref{Stationary_equation_perp}) is nothing but the dynamical equation of the pure system, where the random force acts as an effective thermal noise.
It is worth to note that the correlation of $F(x,\bvec{r}_{\perp};\phi)$ for different field values is irrelevant because, at a specific coordinate $(x,\bvec{r}_{\perp})$, the field $\phi$ feels the random force only once in the forward propagation along $x$-direction.
In other words, $F(x,\bvec{r}_{\perp};\phi)$ can be replaced with a field-independent random force $f(x,\bvec{r}_{\perp})$, whose cumulant is given by
\begin{equation} 
\overline{f(x,\bvec{r}_{\perp}) f(x',\bvec{r}_{\perp}')}=\Delta(0)\delta(x-x')\delta(\bvec{r}_{\perp}-\bvec{r}_{\perp}').
\end{equation}
Therefore, we can conclude that the cross-section of the $D$-dimensional driven disordered system at zero temperature is identical to the corresponding $(D-1)$-dimensional pure system in equilibrium with a temperature $\Delta(0)/(2v)$.

Below, we also provide a formal derivation of the dimensional reduction by using a path integral formalism.
For an arbitrary functional $A[\phi]$, we have
\begin{eqnarray}
A[\phi_{\mathrm{st}}] &=& \int \mathcal{D} \phi A[\phi] \delta[\phi-\phi_{\mathrm{st}}] \nonumber \\
&=& \int \mathcal{D} \phi A[\phi] \delta[v\partial_x \phi - \nabla^2 \phi + U'(\phi) - F(\bvec{r};\phi)] \mathcal{J}[\phi],
\label{A_phi_st}
\end{eqnarray}
where the Jacobian $\mathcal{J}[\phi]$ is given by
\begin{eqnarray}
\mathcal{J}[\phi] = |\mathrm{det}(v\partial_x-\nabla^2 + U''(\phi) - \partial_{\phi} F(\bvec{r};\phi))|.
\label{Jacobian_DR}
\end{eqnarray}
The delta function in Eq.~(\ref{A_phi_st}) can be rewritten in terms of an auxiliary field $\hat{\phi}$ as,
\begin{eqnarray}
\delta[v\partial_x \phi - \nabla^2 \phi + U'(\phi) - F(\bvec{r};\phi)] \nonumber \\
= \int \mathcal{D} \hat{\phi} \exp \biggl[ - \int_r i\hat{\phi} (v\partial_x \phi - \nabla^2 \phi + U'(\phi) - F(\bvec{r};\phi)) \biggr].
\end{eqnarray}
By introducing two anticommuting (Grassmann) fields $\psi$ and $\psi^*$, the Jacobian can be rewritten as
\begin{eqnarray}
\mathcal{J}[\phi] = \int \mathcal{D} \psi^* \mathcal{D} \psi \exp \left[-\int_{r} \psi^* (v\partial_x - \nabla^2 + U''(\phi) - \partial_{\phi} F(\bvec{r};\phi)) \psi \right],
\end{eqnarray}
where we have assumed that the determinant in Eq.~(\ref{Jacobian_DR}) is positive because the stationary state should be stable.
We take the average over the disorder,
\begin{eqnarray}
\overline{A[\phi_{\mathrm{st}}]} = \int \mathcal{D} \phi \mathcal{D} \hat{\phi} \mathcal{D} \psi^* \mathcal{D} \psi A[\phi] \exp(-S[\phi,i\hat{\phi}]-\tilde{S}[\psi,\psi^*]),
\label{DR_A_average}
\end{eqnarray}
\begin{eqnarray}
S[\phi,i\hat{\phi}] = \int_{r} \biggl[i\hat{\phi} (v\partial_x  \phi - \nabla^2 \phi + U'(\phi)) - \frac{1}{2} (i\hat{\phi})^2 \Delta(0) \biggr],
\label{DR_S}
\end{eqnarray}
\begin{eqnarray}
\tilde{S}[\psi,\psi^*] = \int_{r} \psi^* (v\partial_x - \nabla^2 + U''(\phi)) \psi,
\label{DR_S_tilde}
\end{eqnarray}
where we have used $\overline{F(\bvec{r};\phi) \partial_{\phi} F(\bvec{r};\phi)}=0$ and $\overline{\exp(\psi^* \partial_{\phi} F \psi)}=1$.
Note that the value of $\Delta(\phi)$ at $\phi \neq 0$ does not appear in the action, because the correlation of the random force is evaluated for the field at the same spatial point owing to the delta correlated nature of the random force (see Eq.~(\ref{bare_random_force_cumulant})).
As mentioned above, since the higher derivatives in the longitudinal direction $\partial_x^2 \phi$ is irrelevant to the long-distance physics, one can omit it.
If the longitudinal coordinate $x$ is replaced with a fictitious time $\tau$, Eqs.~(\ref{DR_S}) and (\ref{DR_S_tilde}) coincide with the action corresponding to the dynamical equation of the $(D-1)$-dimensional pure system, 
\begin{eqnarray}
v\partial_{\tau} \phi(\tau,\bvec{r}_{\perp}) = \nabla_{\perp}^2 \phi(\tau,\bvec{r}_{\perp}) - U'(\phi(\tau,\bvec{r}_{\perp}))+ f(\tau,\bvec{r}_{\perp}),
\end{eqnarray}
where $\overline{f(\tau,\bvec{r}_{\perp})f(\tau',\bvec{r}_{\perp}')} = \Delta(0) \delta(\tau-\tau') \delta(\bvec{r}_{\perp}-\bvec{r}_{\perp}')$.
If $A[\phi]$ is chosen as $\phi(\bvec{r})\phi(\bvec{0})$, it is concluded that the equal-time correlation function of the driven disordered system is identical to the dynamical correlation function of the $(D-1)$-dimensional pure system.

It is well known that the supersymmetry plays a crucial role in the conventional dimensional reduction.
In thermal equilibrium ($v=0$), the action Eqs.~(\ref{DR_S}) and (\ref{DR_S_tilde}) can be cast into a supersymmetric form, which is invariant with respect to the rotation in a superspace spanned by the $D$-dimensional coordinates and two additional anticommuting coordinates \cite{Parisi-79}.
This supersymmetry leads to the reduction from $D$ to $D-2$ dimensions.
In contrast, for the nonequilibrium case ($v \neq 0$), the supersymmetry is explicitly broken owing to the presence of the driving term $v\partial_x \phi$.
Thus, the mechanism of the dimensional reduction in driven systems is quite different from that in equilibrium and it has nothing to do with the supersymmetry.

Unfortunately, as in equilibrium, the dimensional reduction in driven systems does not always hold.
For example, let us consider the case of the RFIM, $U(\phi)=\lambda(\phi^2-\phi_0^2)^2$ and $\Delta(\phi)=\Delta_0$.
Recall that, in equilibrium ($v=0$), the lower critical dimension of the RFIM is two \cite{Imry-75,Imbrie-84,Bricmont-87}.
We expect that the transverse driving reduces the critical dimensions by one.
In fact, the lower critical dimension of the driven random field XY model is shown to be three, while that of the random field XY model is four \cite{Haga-15,Haga-18}.
Thus, we can conclude that the two-dimensional driven RFIM exhibits long-range order at weak disorder.
However, this contradicts the dimensional reduction because it predicts that the driven RFIM in two dimensions is identical to the pure Ising model in one dimension, which does not exhibit any phase transition.

The failure of the dimensional reduction is a consequence of the fact that there are a large number of stationary states satisfying Eq.~(\ref{Stationary_equation}).
Let $\{ \phi_{\mathrm{st}}^{(\chi)}(\bvec{r}) \}_{\chi=1,...,\mathcal{N}}$ be the set of all stationary states.
Then, the correlation function calculated from Eqs.~(\ref{DR_A_average}), (\ref{DR_S}), and (\ref{DR_S_tilde}) is equal to $C_{\mathrm{uni}}(\bvec{r})$ given by Eq.~(\ref{correlation_eq_uniform}), where ``$\mathrm{sign}(\chi)$'' represents the sign of $\mathrm{det}(v \partial_x + \delta^2 \mathcal{H}/\delta \phi \delta \phi)$ evaluated at $\phi=\phi_{\mathrm{st}}^{(\chi)}$.
Thus, we can conclude that $C_{\mathrm{uni}}(x,\bvec{r}_{\perp})$ is identical to the dynamical correlation function of the $(D-1)$-dimensional pure model by replacing the coordinate $x$ with a fictitious time $\tau$.
However, it is nontrivial whether $C_{\mathrm{uni}}(\bvec{r})$ is identical to the actual correlation function, which contains the average over the stationary states with a nontrivial weight.

It is worth noting that there is some ambiguity in the definition of the correlation function $C(\bvec{r})$ at zero temperature.
In the presence of the thermal noise $\xi$, the probability distribution for the steady state $P_{\mathrm{ss}}[\phi;V]$ is unique, provided that the dynamics of the system is ergodic.
Then, $C(\bvec{r})$ is defined by Eq.~(\ref{Disorder_average_A}) without any ambiguity.
However, at zero temperature or without the thermal noise, $P_{\mathrm{ss}}[\phi;V]$ is ill-defined.
In thermal equilibrium, such a problem does not exist because the probability distribution is written in terms of the Hamiltonian.
Then, $C(\bvec{r})$ at zero temperature is calculated from the ground state.
Note that, for the nonequilibrium case,  the notion of ``ground state'' is meaningless because Eq.~(\ref{General_EM}) cannot be cast into the form of Eq.~(\ref{General_EM_eq}) with any Hamiltonian.
In the following, we propose different versions of $C(\bvec{r})$ for driven disordered systems at zero temperature.
In the first definition, we choose an initial condition $\phi_{\mathrm{i}}(\bvec{r})$ according to the probability distribution function $P_{\mathrm{i}}[\phi;\mu] \sim e^{-(\mu/2)\phi^2}$ and solve Eq.~(\ref{General_EM_zero_temp}) to obtain a stationary state $\phi_{\mathrm{st}}(\bvec{r})$, which is one of the solutions of Eq.~(\ref{Stationary_equation}).
Since there are many stationary states, $\phi_{\mathrm{st}}(\bvec{r})$ depends on the initial condition $\phi_{\mathrm{i}}(\bvec{r})$.
We then define $C_1(\bvec{r})$ by
\begin{equation}
C_1(\bvec{r}) = \lim_{\mu \to 0} \int \mathcal{D}V P_{\mathrm{R}}[V] \langle \phi_{\mathrm{st}}(\bvec{r}) \phi_{\mathrm{st}}(\bvec{0}) \rangle_{\mathrm{i}},
\end{equation}
where $\langle...\rangle_{\mathrm{i}}$ represents the average over the probability distribution function of the initial condition $P_{\mathrm{i}}[\phi;\mu]$.
In the second definition, we switch to the moving frame, $\bvec{r}'=\bvec{r}-\bvec{v}t$ and $\phi(\bvec{r},t)=\phi'(\bvec{r}',t)$. Eq.~(\ref{General_EM_zero_temp}) is then rewritten as
\begin{equation}
\partial_t \phi'(\bvec{r}',t) = \nabla^2 \phi'(\bvec{r}',t) - U'(\phi'(\bvec{r}',t))+ F(\bvec{r}'+\bvec{v}t;\phi'(\bvec{r}',t)),
\label{General_EM_zero_temp_moving}
\end{equation}
where the last term of the right-hand side is a uniformly moving random force.
We assume free boundary conditions, not periodic boundary conditions, to prevent the system from feeling the same disorder periodically. 
The new random force is continuously generated in the front boundary of the system and moves with the velocity $-\bvec{v}$.
Let $\phi(\bvec{r},t)$ be a solution of Eq.~(\ref{General_EM_zero_temp_moving}) with the above boundary conditions.
We then define $C_2(\bvec{r})$ by
\begin{equation}
C_2(\bvec{r}) = \lim_{\tau \to \infty} \frac{1}{\tau} \int_0^{\tau} \phi(\bvec{r},t) \phi(\bvec{0},t) dt.
\end{equation}
Although it is nontrivial whether $C_1(\bvec{r})$ coincides with $C_2(\bvec{r})$, we assume it in this study.
We emphasize that the ambiguity mentioned above comes from the choice of the weight used in the average over the multiple stationary states.
If there is only a single stationary state satisfying Eq.~(\ref{Stationary_equation}), the correlation function can be uniquely defined.
In other words, this ambiguity in averaged quantities for driven systems without thermal noise has the same origin as the breakdown of the dimensional reduction argument.

\section{NP-FRG formalism}
\label{sec:NP-FRG formalism}

The standard perturbative approach leads to the dimensional reduction, which is incorrect in general.
To describe the breakdown of the dimensional reduction, the functional renormalization group (FRG) theory has been developed for disordered systems in thermal equilibrium \cite{Fisher-85-2,Fisher-86,Balents-93,Giamarchi-95,Feldman-02,
LeDoussal-03,LeDoussal-04}.
In this approach, one considers the evolution of the whole functional form of the random force cumulant $\Delta(\phi)$ when the high-energy modes are successively integrated out.
The long-distance physics of the system is then controlled by the corresponding fixed point function $\Delta_*(\phi)$.
For some cases, this fixed point has a linear cusp as a function of the field, $\Delta_*(\phi) \simeq \Delta_*(0) + \Delta_*'(0^+)|\phi|$ near $\phi=0$.
The generation of the cusp in the renormalized disorder cumulant is a signature of the presence of many local minima in the energy landscape, and it leads to the breakdown of the dimensional reduction.
In the following, we will apply the FRG approach to our nonequilibrium models and demonstrate how and when the dimensional reduction breaks down.

\subsection{Effective action}

We consider a multi-component elastic manifold transversely driven in a random potential, which is defined by Eqs.~(\ref{General_Hamiltonian}) and (\ref{General_EM}) with $U(\phi)=0$.
The field $\phi$ is now an $N$-component vector field $\bvec{\phi}={}^t(\phi^1,...,\phi^N)$ and the second cumulant of the random force is given by 
\begin{equation}
\overline{F^{\alpha}(\bvec{r};\bvec{\phi})F^{\beta}(\bvec{r}';\bvec{\phi}')}=\Delta^{\alpha \beta}(\bvec{\phi}-\bvec{\phi}')\delta(\bvec{r}-\bvec{r}').
\end{equation}
We call this model the driven random manifold (DRM).
If we define the second cumulant of the random potential $R(\rho)$ by
\begin{equation}
\overline{V(\bvec{r};\bvec{\phi})V(\bvec{r}';\bvec{\phi}')}=R \biggl( \frac{(\bvec{\phi}-\bvec{\phi}')^2}{2} \biggr) \delta(\bvec{r}-\bvec{r}'),
\end{equation}
then the second cumulant of the random force can be written as
\begin{eqnarray}
\Delta^{\alpha \beta}(\bvec{\phi}) = - \partial_{\alpha} \partial_{\beta} R(\rho),
\label{Delta_R}
\end{eqnarray}
where $\partial_{\alpha} = \partial/\partial \phi^{\alpha}$ and $\rho=|\bvec{\phi}|^2/2$.
We consider two types of disorder, the short-range correlated disorder,
\begin{eqnarray}
R(\rho) \sim \exp(-\rho),
\label{R_short_range}
\end{eqnarray}
and the long-range correlated disorder,
\begin{eqnarray}
R(\rho) \sim \frac{\rho^{1-\gamma}}{\gamma-1}.
\label{R_long_range}
\end{eqnarray}
A quantity of interest is the roughness exponent for the transverse direction $\zeta_{\perp}$, which is defined by 
\begin{equation}
\overline{\langle |\bvec{\phi}(x,\bvec{r}_{\perp})-\bvec{\phi}(x,\bvec{r}_{\perp}')|^2 \rangle} \sim |\bvec{r}_{\perp}-\bvec{r}_{\perp}'|^{2\zeta_{\perp}}.
\end{equation}
The dimensional reduction predicts that 
\begin{equation}
\zeta_{\perp,\mathrm{DR}} =\frac{3-D}{2}.
\end{equation}

To derive the FRG equation for the disorder cumulant $\Delta^{\alpha \beta}(\bvec{\phi})$, we employ the so-called nonperturbative FRG (NP-FRG) formalism developed in Refs.~\cite{Tarjus-08,Tissier-08,Tissier-12-1,Tissier-12-2}.
First, the equation of motion (\ref{General_EM}) is rewritten in the form of the field theoretical action.
We introduce an $n$-replicated system with the same disorder,
\begin{equation}
(\partial_t + v \partial_x) \phi_a^{\alpha}(\bvec{r},t) = \nabla^2 \phi_a^{\alpha}(\bvec{r},t) + F^{\alpha}(\bvec{r};\bvec{\phi}_a(\bvec{r},t))+\xi_a^{\alpha}(\bvec{r},t),
\label{replicated_EM}
\end{equation}
where the superscript of Greek alphabet ($\alpha,\beta,...=1,...,N$) represents the index of the field component and the subscript of Roman alphabet ($a,b,...=1,...,n$) represents the replica index.
The thermal noise satisfies
\begin{equation}
\langle \xi_a^{\alpha}(\bvec{r},t) \xi_b^{\beta}(\bvec{r}',t') \rangle =2 T \delta^{\alpha \beta} \delta_{ab} \delta(\bvec{r}-\bvec{r'}) \delta(t-t').
\end{equation}
For an arbitrary functional $A[\{\bvec{\phi}_a\}]$, its average over the thermal noise is written as
\begin{eqnarray}
\langle A[\{\bvec{\phi}_a\}] \rangle = \int \mathcal{D} \xi P[\xi] \int \mathcal{D} \bvec{\phi} \delta(\bvec{\phi}_a-\bvec{\phi}_a[\xi]) A[\{\bvec{\phi}_a\}],
\end{eqnarray}
where $\bvec{\phi}_a[\xi]$ is the solution of Eq.~(\ref{replicated_EM}) for a realization of the noise $\xi_a$.
This average can be calculated as
\begin{eqnarray}
\langle A[\{\bvec{\phi}_a\}] \rangle &=& \int \mathcal{D} \xi P[\xi] \int \mathcal{D} \bvec{\phi} A[\{\bvec{\phi}_a\}]  \delta \bigl[ (\partial_t + v \partial_x) \phi_a^{\alpha} \nonumber \\
&&- \nabla^2 \phi_a^{\alpha} - F^{\alpha}(\bvec{r};\bvec{\phi}_a) - \xi_a^{\alpha} \: \bigr] \nonumber \\
&=& \int \mathcal{D} \xi P[\xi] \int \mathcal{D} \bvec{\phi} \mathcal{D} \bvec{\hat{\phi}} A[\{\bvec{\phi}_a\}]  \exp \biggl[ - \sum_a \sum_{\alpha} \nonumber \\
&&\int_{rt} i \hat{\phi}_a^{\alpha} \bigl\{ (\partial_t + v \partial_x) \phi_a^{\alpha} - \nabla^2 \phi_a^{\alpha} - F^{\alpha}(\bvec{r};\bvec{\phi}_a) - \xi_a^{\alpha} \bigr\} \biggr] \nonumber \\
&=& \int \mathcal{D} \bvec{\phi} \mathcal{D} \bvec{\hat{\phi}} A[\{\bvec{\phi}_a\}] \exp \biggl[ - \sum_a \sum_{\alpha} \int_{rt} i \hat{\phi}_a^{\alpha} \bigl\{ (\partial_t + v \partial_x) \phi_a^{\alpha} \nonumber \\
&&- T i \hat{\phi}_a^{\alpha}  - \nabla^2 \phi_a^{\alpha} - F^{\alpha}(\bvec{r};\bvec{\phi}_a)  \bigr\} \biggr],
\end{eqnarray}
where the Jacobian associated with the delta function can be set to unity \cite{Canet-11}.
We next take the average over the disorder $F^{\alpha}(\bvec{r};\bvec{\phi}_a)$,
\begin{eqnarray}
\overline{\langle A[\{\bvec{\phi}_a\}] \rangle} = \int \mathcal{D} \bvec{\phi} \mathcal{D} \bvec{\hat{\phi}} A[\{\bvec{\phi}_a\}] \exp \bigl( -S[\{ \bvec{\phi}_a, \bvec{\hat{\phi}}_a \}] \bigr),
\end{eqnarray}
where the disorder averaged action $S[\{ \bvec{\phi}_a, \bvec{\hat{\phi}}_a \}]$ is given by
\begin{eqnarray}
S[\{ \bvec{\phi}_a, \bvec{\hat{\phi}}_a \}] &=& \sum_a \int_{rt} i\bvec{\hat{\phi}}_a \cdot \bigl[ \partial_t \bvec{\phi}_a - T i\bvec{\hat{\phi}}_a + v \partial_x \bvec{\phi}_a - \nabla^2 \bvec{\phi}_a \bigr] \nonumber \\
&&- \frac{1}{2} \sum_{a,b} \int_{rtt'} i\hat{\phi}_{a,rt}^{\alpha} i\hat{\phi}_{b,rt'}^{\beta} \Delta^{\alpha \beta}( \bvec{\phi}_{a,rt} - \bvec{\phi}_{b,rt'} ),
\label{DRM_bare_action}
\end{eqnarray}
where the summation over repeated indices $\alpha$ and $\beta$ is assumed.
In the following, we omit the imaginary unit $i$ in $i\hat{\phi}$ for simplicity, and we write $\Phi_a = {}^t( \bvec{\phi}_a, \bvec{\hat{\phi}}_a )$.

By introducing a source field $J_a={}^t(\bvec{j}_a,\bvec{\hat{j}}_a)$, the generating functional $W[\{J_a\}]$ is defined by
\begin{equation}
e^{W[\{J_a\}]}=\int \mathcal{D} \Phi \exp \biggl[ -S[\{ \Phi_a \}] + \sum_a \int_{rt}  {}^tJ_a \cdot \Phi_a \biggr].
\end{equation}
The effective action is then given by a Legendre transform, 
\begin{equation}
\Gamma[\{ \Psi_a \}] = -W[\{ J_a \}] + \sum_a \int_{rt} {}^tJ_a \cdot \Psi_a,
\end{equation}
where $\Psi_a ={}^t(\bvec{\psi}_a, \bvec{\hat{\psi}}_a)$ and $J_a$ are related by 
\begin{equation}
\Psi_a = \frac{\delta}{\delta J_a} W[\{J_a \}].
\end{equation}

The basic concept of the NP-FRG formalism is to construct the scale-dependent effective action $\Gamma_k[\{ \Psi_a \}]$, which includes only high-energy modes with momenta larger than the running scale $k$.
As $k$ goes from the cutoff $\Lambda$ to zero, $\Gamma_k[\{ \Psi_a \}]$ interpolates between the bare action $S[\{ \Psi_a \}]$ and the full effective action $\Gamma[\{ \Psi_a \}]$.
The effective action $\Gamma_k[\{ \Psi_a \}]$ is defined as follows.
To suppress the contribution from the low-energy modes, a mass-like quadratic term is added to the bare action,
\begin{equation}
\Delta S_k[\{ \Phi_a \}] = \frac{1}{2} \sum_a \int_{q} {}^t\Phi_a(q) \: \mathbf{R}_k(\mathbf{q}) \: \Phi_a(-q),
\label{Delta_S_k}
\end{equation}
where we have used notations $q=(\mathbf{q},\omega)$ and $\int_q =\int d^D \mathbf{q} d \omega/(2\pi)^{D+1}$.
A frequency-independent $2N \times 2N$ matrix $\mathbf{R}_k(\mathbf{q})$ is given by
\begin{eqnarray}
\mathbf{R}_k(\mathbf{q}) = R_k(\mathbf{q}) \left(
\begin{array}{ccc}
0 & 1 \\
1 & 0
\end{array}
\right) \otimes \mathbf{I}_N,
\end{eqnarray}
where $\mathbf{I}_N$ is the $N \times N$ unit matrix, which acts on the field component index.
$R_k(\mathbf{q})$ is a cutoff function, which has a constant value proportional to $k^2$ for $q \ll k$ and rapidly decreases for $q > k$.
The generating functional $W_k[\{ J_a \}]$ with the running scale $k$ is defined by 
\begin{equation}
e^{W_k[\{ J_a \}]}=\int \mathcal{D} \Phi \exp \biggl[ -S[\{ \Phi_a \}] - \Delta S_k[\{ \Phi_a \}] + \sum_a \int_{rt}  {}^tJ_a \cdot \Phi_a \biggr].
\end{equation}
The effective action then reads 
\begin{equation}
\Gamma_k[\{ \Psi_a \}] = -W_k[\{ J_a \}] + \sum_a \int_{rt} {}^tJ_a \cdot \Psi_a - \Delta S_k[\{ \Psi_a \}],
\end{equation}
where $\Psi_a$ and $J_a$ are related by 
\begin{equation}
\Psi_a = \frac{\delta}{\delta J_a} W_k[\{J_a \}].
\end{equation}

\subsection{Exact flow equation}

The flow of $\Gamma_k$ is described by the Wetterich equation \cite{Wetterich-00},
\begin{equation}
\partial_k \Gamma_k = \frac{1}{2} \mathrm{Tr} \: \partial_k  \hat{\mathbf{R}}_k(\mathbf{q}) \Bigl[ \Gamma_k^{(2)} + \hat{\mathbf{R}}_k(\mathbf{q}) \Bigr]^{-1},
\label{Wetterich_equation}
\end{equation}
where $\Gamma_k^{(2)}$ is the second functional derivative of $\Gamma_k$ and ``$\mathrm{Tr}$'' represents an integration over momentum and frequency as well as a sum over replica indices, field component indices, and the two conjugate fields $\{ \psi, \hat{\psi} \}$.
We have introduced a $2nN \times 2nN$ matrix
\begin{eqnarray}
\hat{\bvec{\mathrm{R}}}_k(\mathbf{q}) = \bvec{\mathrm{R}}_k(\mathbf{q}) \otimes \mathbf{I}_n,
\end{eqnarray}
where $\mathbf{I}_n$ is the $n \times n$ unit matrix, which acts on the replica index.

Since $\Gamma_k$ reaches the bare action Eq.~(\ref{DRM_bare_action}) in the limit $k \to \infty$, $\Gamma_k$ is also expanded by increasing the number of free replica sums as
\begin{equation}
\Gamma_k[\{ \Psi_a \}] = \sum_{p=1}^{\infty} \sum_{a_1,...,a_p} \frac{(-1)^{p-1}}{p!} \Gamma_{p,k}[\Psi_{a_1},...,\Psi_{a_p}],
\label{Gamma_expansion}
\end{equation}
where $\Gamma_{p,k}$ corresponds to the $p$-th cumulant of the disorder.
Although the bare random force is chosen as Gaussian, the higher order cumulants can be generated along the RG flow.
By substituting Eq.~(\ref{Gamma_expansion}) into Eq.~(\ref{Wetterich_equation}), we obtain the exact flow equations for $\Gamma_{p,k}$.
To express these in a compact form, we define the one-replica propagator with the infrared cutoff, 
\begin{equation}
\mathrm{P}_k[\Psi] = \Bigl[ \Gamma_{1,k}^{(2)}[\Psi] + \mathbf{R}_k(\mathbf{q}) \Bigr]^{-1}.
\label{one_replica_propagator}
\end{equation}
To calculate the inverse of $\Gamma_k^{(2)} + \hat{\mathbf{R}}_k$ with respect to the replica indices, we rewrite it as
\begin{eqnarray}
\bigl( \Gamma_k^{(2)} + \hat{\mathbf{R}}_k \bigr)_{ab} = \mathrm{P}_k[\Psi_a]^{-1} \delta_{ab} - A[\Psi_a] \delta_{ab} - B[\Psi_a,\Psi_b].
\label{Gamma(2)_R}
\end{eqnarray}
We explicitly write only the replica index, and omit the indices of momentum, frequency, field component, and two conjugate fields.
$A[\Psi_a]$ and $B[\Psi_a,\Psi_b]$ can be also expanded by increasing the number of free replica sums,
\begin{eqnarray}
A[\Psi_a] = \sum_c A^{[1]}[\Psi_a|\Psi_c] + \frac{1}{2} \sum_{c,d} A^{[2]}[\Psi_a|\Psi_c,\Psi_d] +...,
\label{A_expansion}
\end{eqnarray}
\begin{eqnarray}
B[\Psi_a,\Psi_b] = B^{[0]}[\Psi_a,\Psi_b] + \sum_c B^{[1]}[\Psi_a,\Psi_b|\Psi_c] + ...,
\label{B_expansion}
\end{eqnarray}
where the vertical bar in each term $A^{[p]}[\Psi_a|\Psi_{c_1},...,\Psi_{c_p}]$ is introduced to distinguish between the ``explicit'' index $a$ and the dummy indices $c_1,...,c_p$, which run from $1$ to $n$ as the summation is taken.
In the following, we use simplified notations such as,
\begin{eqnarray}
\Gamma_{2,k}^{(20)}[\Psi_1,\Psi_2] = \frac{\delta^2 \Gamma_{2,k}[\Psi_1,\Psi_2]}{\delta \Psi_1 \delta \Psi_1}, \nonumber \\
\Gamma_{3,k}^{(110)}[\Psi_1,\Psi_2,\Psi_3] = \frac{\delta^2 \Gamma_{3,k}[\Psi_1,\Psi_2,\Psi_3]}{\delta \Psi_1 \delta \Psi_2}.
\end{eqnarray}
From Eq.~(\ref{Gamma_expansion}), the first two terms of $A^{[p]}$ and $B^{[p]}$ are written as
\begin{eqnarray}
A^{[1]}[\Psi_a|\Psi_c] = \Gamma_{2,k}^{(20)}[\Psi_a,\Psi_c], \nonumber \\
A^{[2]}[\Psi_a|\Psi_c,\Psi_d] = - \Gamma_{3,k}^{(200)}[\Psi_a,\Psi_c,\Psi_d],
\label{A_Gamma}
\end{eqnarray}
and
\begin{eqnarray}
B^{[0]}[\Psi_a,\Psi_b] = \Gamma_{2,k}^{(11)}[\Psi_a,\Psi_b], \nonumber \\
B^{[1]}[\Psi_a,\Psi_b|\Psi_c] = - \Gamma_{3,k}^{(110)}[\Psi_a,\Psi_b,\Psi_c].
\label{B_Gamma}
\end{eqnarray}
The inverse of Eq.~(\ref{Gamma(2)_R}) is expanded as
\begin{eqnarray}
\bigl( \Gamma_k^{(2)} + \hat{\mathbf{R}}_k \bigr)^{-1}_{ab} &=& \mathrm{P}_k[\Psi_a] \delta_{ab} + \mathrm{P}_k[\Psi_a]\bigl( A[\Psi_a] \delta_{ab}+B[\Psi_a,\Psi_b]\bigr)\mathrm{P}_k[\Psi_b] \nonumber \\
&&+ \sum_c \mathrm{P}_k[\Psi_a]\bigl( A[\Psi_a] \delta_{ac}+B[\Psi_a,\Psi_c] \bigr)\mathrm{P}_k[\Psi_c] \nonumber \\
&&\times \bigl( A[\Psi_c] \delta_{cb}+B[\Psi_c,\Psi_b] \bigr)\mathrm{P}_k[\Psi_b] +....
\end{eqnarray}
By substituting Eqs.~(\ref{A_expansion}) and (\ref{B_expansion}) into the above equation and taking the trace,
\begin{eqnarray}
\sum_a \bigl( \Gamma_k^{(2)} + \hat{\mathbf{R}}_k \bigr)^{-1}_{aa} = \sum_{p=1}^{\infty} \sum_{a_1,...,a_p} \frac{(-1)^{p-1}}{p!} Q_{p}[\Psi_{a_1},...,\Psi_{a_p}].
\end{eqnarray}
The first two terms $Q_1$ and $Q_2$ are given by
\begin{eqnarray}
Q_1[\Psi_a] = \mathrm{P}_k[\Psi_a] + \mathrm{P}_k[\Psi_a] B^{[0]}[\Psi_a,\Psi_a] \mathrm{P}_k[\Psi_a],
\label{Q_1}
\end{eqnarray}
\begin{eqnarray}
Q_2[\Psi_a,\Psi_b] &=& \mathrm{P}_k[\Psi_a] \bigl\{ A^{[1]}[\Psi_a|\Psi_b]  \nonumber \\
&&+ B^{[0]}[\Psi_a,\Psi_b] \mathrm{P}_k[\Psi_b] B^{[0]}[\Psi_b,\Psi_a]  \nonumber \\
&&+ A^{[1]}[\Psi_a|\Psi_b] \mathrm{P}_k[\Psi_a] B^{[0]}[\Psi_a,\Psi_a] \nonumber \\
&&+ B^{[0]}[\Psi_a,\Psi_a] \mathrm{P}_k[\Psi_a] A^{[1]}[\Psi_a|\Psi_b] \nonumber \\
&&+ B^{[1]}[\Psi_a,\Psi_a|\Psi_b] \bigr\} \mathrm{P}_k[\Psi_a] + \mathrm{perm}(\Psi_a,\Psi_b),
\label{Q_2}
\end{eqnarray}
where $\mathrm{perm}(\Psi_a,\Psi_b)$ is the expression obtained by permuting $\Psi_a$ and $\Psi_b$.
From Eqs.~(\ref{A_Gamma}), (\ref{B_Gamma}), (\ref{Q_1}), and (\ref{Q_2}), we obtain the exact flow equations for $\Gamma_{1,k}[ \Psi ]$ and $\Gamma_{2,k}[ \Psi_1, \Psi_2 ]$,
\begin{eqnarray}
\partial_k \Gamma_{1,k}[ \Psi ] = \frac{1}{2} \mathrm{tr} \int_{q} \partial_k \bvec{\mathrm{R}}_k(\mathbf{q}) \Bigl[ \mathrm{P}_k[\Psi] + \mathrm{P}_k[\Psi] \Gamma_{2,k}^{(11)}[\Psi,\Psi] \mathrm{P}_k[\Psi] \Bigr]_{q,-q},
\label{Exact_flow_Gamma_1}
\end{eqnarray}
\begin{eqnarray}
\partial_k \Gamma_{2,k}[ \Psi_1, \Psi_2 ] &=& -\frac{1}{2} \mathrm{tr} \int_{q} \partial_k \bvec{\mathrm{R}}_k(\mathbf{q}) \Bigl[ \mathrm{P}_k[\Psi_1] \Bigl\{
\Gamma_{2,k}^{(20)}[\Psi_1,\Psi_2] \nonumber \\
&&+ 2 \Gamma_{2,k}^{(20)}[\Psi_1,\Psi_2] \mathrm{P}_k[\Psi_1] \Gamma_{2,k}^{(11)}[\Psi_1,\Psi_1] \nonumber \\
&&+ \Gamma_{2,k}^{(11)}[\Psi_1,\Psi_2] \mathrm{P}_k[\Psi_2] \Gamma_{2,k}^{(11)}[\Psi_2,\Psi_1] \nonumber \\
&&- \Gamma_{3,k}^{(110)}[\Psi_1,\Psi_1,\Psi_2] \Bigr\} \mathrm{P}_k[\Psi_1] +\mathrm{perm}(\Psi_1,\Psi_2) \Bigr]_{q,-q},
\label{Exact_flow_Gamma_2}
\end{eqnarray}
where ``$\mathrm{tr}$'' in Eqs.~(\ref{Exact_flow_Gamma_1}) and (\ref{Exact_flow_Gamma_2}) represents a sum over the indices of the field component and the two conjugate fields $\{ \psi, \hat{\psi} \}$.

\subsection{Flow equation for the disorder cumulant}

To solve the flow equations (\ref{Exact_flow_Gamma_1}) and (\ref{Exact_flow_Gamma_2}), we are required to introduce approximations for the functional forms of $\Gamma_{p,k}$.
We employ the following expression for the one-replica part,
\begin{equation}
\Gamma_{1,k}[\Psi] = \int_{rt} \bvec{\hat{\psi}} \cdot \bigl[ X_k (\partial_t \bvec{\psi} - T_k \bvec{\hat{\psi}}) + v \partial_x \bvec{\psi} - \nabla^2 \bvec{\psi} \bigr],
\label{Gamma_1}
\end{equation}
where $X_k$ and $T_k$ are the scale-dependent relaxation coefficient and temperature, respectively.
For the multi-replica part,
\begin{equation}
\Gamma_{p,k}[\Psi_1,...,\Psi_p] = \sum_{\alpha_1,...,\alpha_p} \int_{rt_1...t_p} \hat{\psi}_{1,rt_1}^{\alpha_1} ... \hat{\psi}_{p,rt_p}^{\alpha_p} \Delta_{p,k}^{\alpha_1...\alpha_p}( \bvec{\psi}_{1,rt_1}, ... , \bvec{\psi}_{p,rt_p} ),
\label{Gamma_n}
\end{equation} 
where $\Delta_{p,k}^{\alpha_1...\alpha_p}(\bvec{\psi}_1,...,\bvec{\psi}_p)$ is the $p$-th cumulant of the renormalized random force, which is symmetric with respect to the permutation of $\bvec{\psi}_s$ and $\alpha_s$, and satisfies
\begin{equation}
\Delta_{p,k}^{\alpha_1...\alpha_p}(\bvec{\psi}_1+\bvec{\lambda},...,\bvec{\psi}_p+\bvec{\lambda}) = \Delta_{p,k}^{\alpha_1...\alpha_p}(\bvec{\psi}_1,...,\bvec{\psi}_p),
\end{equation} 
for an arbitrary vector $\bvec{\lambda}$.
Within Eqs.~(\ref{Gamma_1}) and (\ref{Gamma_n}), the coefficient of $\nabla^2 \bvec{\psi}$ and the driving velocity $v$ are not renormalized because $\partial_k \Gamma_{1,k}^{(2)}$ is independent of the external momenta.
In contrast, the relaxation coefficient $X_k$ and temperature $T_k$ can be renormalized.
From the functional form Eq.~(\ref{Gamma_n}), the flow equation for $\Delta_p$ is obtained from
\begin{equation}
\partial_k \Delta_{p,k}^{\alpha_1...\alpha_p}(\bvec{\psi}_1,...,\bvec{\psi}_p) = \frac{\delta^p}{\delta \hat{\psi}_1^{\alpha_1}... \delta \hat{\psi}_p^{\alpha_p} } \partial_k \Gamma_{p,k}[\Psi_1,...,\Psi_p],
\label{RGeq_Delta-0}
\end{equation}
where the functional derivative is evaluated at a uniform field configuration: $ \bvec{\psi}_{1,rt} \equiv \bvec{\psi}_1,...,\bvec{\psi}_{p,rt} \equiv \bvec{\psi}_p $ and  $ \bvec{\hat{\psi}}_{1,rt} \equiv 0,.., \bvec{\hat{\psi}}_{p,rt} \equiv 0 $.
In the following, we omit the subscript ``$k$'' in $\Delta$.

The one-replica propagator Eq.~(\ref{one_replica_propagator}) has the indices of the field component and the conjugate fields $\{ \psi, \hat{\psi} \}$.
Thus, its elements are written as $P_{ij}^{\alpha \beta}(\mathbf{q},\omega)=P_{ij}(\mathbf{q},\omega)\delta^{\alpha \beta}$, where $\alpha, \beta=1,...,N$ denote the field component indices and $i,j=1,2$ denote the two conjugate fields.
From Eq.~(\ref{Gamma_1}), $P_{ij}(\mathbf{q},\omega)$ is given by
\begin{eqnarray}
P_{11}(\mathbf{q},\omega) &=& \frac{2 X_k T_k}{D(\mathbf{q},\omega)}, \nonumber \\
P_{12}(\mathbf{q},\omega) &=& \frac{M(\mathbf{q})-i(X_k \omega - q_x v)}{D(\mathbf{q},\omega)}, \nonumber \\
P_{21}(\mathbf{q},\omega) &=& \frac{M(\mathbf{q})+i(X_k \omega - q_x v)}{D(\mathbf{q},\omega)}, \nonumber \\
P_{22}(\mathbf{q},\omega) &=& 0,
\label{propagator_component}
\end{eqnarray}
where $M(\mathbf{q}) = |\mathbf{q}|^2 + R_k(\mathbf{q})$ and $D(\mathbf{q},\omega)=M(\mathbf{q})^2 +(X_k \omega - q_x v)^2$.
We also used simplified notations such as $P_{12}(\mathbf{q})=P_{12}(\mathbf{q},\omega=0)$ and $D(\mathbf{q})=D(\mathbf{q},\omega=0)$.

We introduce a new renormalization parameter $l=-\ln (k/\Lambda)$, and then $\partial_l=-k \partial_k$.
From Eq.~(\ref{Exact_flow_Gamma_2}), the flow equation for $\Delta_2$ is given as follows:
\begin{eqnarray}
\partial_l \Delta_2^{\mu \nu}(\bvec{\psi}_a,\bvec{\psi}_b) &=& -\frac{1}{2} \int_{\mathbf{q}} \partial_l R_k(\mathbf{q}) \Bigl[ (\mathrm{A}) + (\mathrm{B}-1) + (\mathrm{B}-2) \nonumber \\
&& + (\mathrm{B}-3) + (\mathrm{C}) + \mathrm{perm}(\bvec{\psi}_a, \bvec{\psi}_b) \Bigr] ,
\label{RG_Delta2_dimensionfull-1}
\end{eqnarray}
\begin{eqnarray}
(\mathrm{A}) &=& \partial_{\psi_1^{\alpha}}^2 \Delta_2^{\mu \nu}(\bvec{\psi}_a,\bvec{\psi}_b) \int_{\omega} [ P_{21}(\mathbf{q};\omega)P_{11}(\mathbf{q};\omega)+P_{11}(\mathbf{q};\omega)P_{12}(\mathbf{q};\omega) ] \nonumber \\
&=& T_k \partial_{\psi_1^{\alpha}}^2 \Delta_2^{\mu \nu}(\bvec{\psi}_a,\bvec{\psi}_b) M(\mathbf{q})^{-2}, \nonumber 
\end{eqnarray}
\begin{eqnarray}
(\mathrm{B}-1) &=& \partial_{\psi_1^{\alpha}} \partial_{\psi_1^{\beta}} \Delta_2^{\mu \nu}(\bvec{\psi}_a,\bvec{\psi}_b) \Delta_2^{\beta \alpha}(\bvec{\psi}_a,\bvec{\psi}_a) \nonumber \\
&&\times [ P_{21}(\mathbf{q})^2 P_{12}(\mathbf{q}) + P_{21}(\mathbf{q}) P_{12}(\mathbf{q})^2 ], \nonumber
\end{eqnarray}
\begin{eqnarray}
(\mathrm{B}-2) &=& \partial_{\psi_1^{\alpha}} \partial_{\psi_2^{\beta}} \Delta_2^{\mu \nu}(\bvec{\psi}_a,\bvec{\psi}_b) \Delta_2^{\beta \alpha}(\bvec{\psi}_b,\bvec{\psi}_a) \nonumber \\
&&\times [ P_{21}(\mathbf{q})^2 P_{12}(\mathbf{q}) + P_{21}(\mathbf{q}) P_{12}(\mathbf{q})^2 ], \nonumber
\end{eqnarray}
\begin{eqnarray}
(\mathrm{B}-3) &=& \partial_{\psi_1^{\alpha}} \Delta_2^{\mu \beta}(\bvec{\psi}_a,\bvec{\psi}_b) \partial_{\psi_1^{\beta}} \Delta_2^{\nu \alpha}(\bvec{\psi}_b,\bvec{\psi}_a) \nonumber \\
&&\times [ P_{21}(\mathbf{q})^3 + P_{12}(\mathbf{q})^3 ], \nonumber  
\end{eqnarray}
\begin{eqnarray}
(\mathrm{C}) = - \partial_{\psi_1^{\alpha}} \Delta_3^{\mu \alpha \nu}(\bvec{\psi}_a,\bvec{\psi}_a,\bvec{\psi}_b) \:  [ P_{21}(\mathbf{q})^2 + P_{12}(\mathbf{q})^2 ], \nonumber 
\end{eqnarray}
where the summation over the repeated indices $\alpha,\beta$ is assumed.
The detailed derivation of Eq.~(\ref{RG_Delta2_dimensionfull-1}) is presented in Appendix.

It is convenient to introduce the following integrals:
\begin{eqnarray}
I_n = -\frac{1}{2} \int_{\mathbf{q}} \partial_l R_k(\mathbf{q}) M(\mathbf{q})^{-n-1},
\label{def_I}
\end{eqnarray}
\begin{eqnarray}
L_n^{-} = -\frac{1}{2} \int_{\mathbf{q}} \partial_l R_k(\mathbf{q}) [ n P_{21}(\mathbf{q})^{n+1} + n P_{12}(\mathbf{q})^{n+1} ],
\label{def_L-}
\end{eqnarray}
\begin{eqnarray}
L_n^{+} = -\frac{1}{2} \int_{\mathbf{q}} \partial_l R_k(\mathbf{q}) \sum_{j=1}^n  2 P_{21}(\mathbf{q})^{n+1-j} P_{12}(\mathbf{q})^j.
\label{def_L+}
\end{eqnarray}
We also define
\begin{eqnarray}
\Delta^{\mu \nu}(\bvec{\psi}_a-\bvec{\psi}_b) &=& \Delta_2^{\mu \nu}(\bvec{\psi}_a,\bvec{\psi}_b), \nonumber \\
\Delta_3^{\mu \nu}(\bvec{\psi}_a-\bvec{\psi}_b) &=& \frac{1}{2} [ \partial_{\psi_1^{\alpha}} \Delta_3^{\mu \alpha \nu} ( \bvec{\psi}_a,\bvec{\psi}_a,\bvec{\psi}_b) + \partial_{\psi_1^{\alpha}} \Delta_3^{\mu \alpha \nu}(\bvec{\psi}_b,\bvec{\psi}_b,\bvec{\psi}_a) ],
\end{eqnarray}
which satisfy $\Delta^{\mu \nu}(-\bvec{\psi})=\Delta^{\mu \nu}(\bvec{\psi})$ and $\Delta_3^{\mu \nu}(-\bvec{\psi})=\Delta_3^{\mu \nu}(\bvec{\psi})$.
Equation (\ref{RG_Delta2_dimensionfull-1}) can be rewritten in the following compact form:
\begin{eqnarray}
\partial_l \Delta^{\mu \nu}(\bvec{\psi}) &=& 2T_k \partial_{\alpha}^2 \Delta^{\mu \nu}(\bvec{\psi}) I_1 + \partial_{\alpha} \partial_{\beta} \Delta^{\mu \nu}(\bvec{\psi}) (\Delta^{\beta \alpha}(\bvec{0})-\Delta^{\beta \alpha}(\bvec{\psi})) L_2^{+} \nonumber \\
&&- \partial_{\alpha} \Delta^{\mu \beta}(\bvec{\psi}) \partial_{\beta} \Delta^{\nu \alpha}(\bvec{\psi}) L_2^{-} -2\Delta_3^{\mu \nu}(\bvec{\psi}) L_1^{-},
\label{RG_Delta}
\end{eqnarray}
where $\partial_{\alpha} = \partial/\partial \psi^{\alpha}$.
In the following, we consider the zero-temperature case $T=T_k=0$.
Note that Eq.~(\ref{RG_Delta}) is not closed due to the presence of the third cumulant $\Delta_3$.

\subsubsection{Equilibrium case}

We first consider the equilibrium cases ($v=0$). 
It is convenient that the momentum $\mathbf{q}$ is measured in units of the running scale $k$,
\begin{equation}
y= \frac{|\mathbf{q}|^2}{k^2}.
\end{equation}
The cutoff function $R_k(\mathbf{q})$ is written as
\begin{eqnarray}
R_k(\mathbf{q}) = k^2 r(y).
\label{cutoff_function_eq}
\end{eqnarray}
In the main calculations, we employ the optimized cutoff function \cite{Litim-00},
\begin{eqnarray}
r(y) = (1-y) \Theta(1-y),
\label{optimized_cutoff_function}
\end{eqnarray}
where $\Theta(x)$ is the step function; $\Theta(x)=1$ for $x>0$ and $\Theta(x)=0$ for $x<0$.
By using this cutoff function, the integrals in Eqs.~(\ref{def_I})--(\ref{def_L+}) are calculated as
\begin{eqnarray}
L_n^- = L_n^+ = 2n I_n = 2n k^{D-2n} \frac{4}{D} A_D,
\end{eqnarray}
where ${A_D}^{-1}=2^{D+1} \pi^{D/2} \Gamma(D/2)$.

To obtain the fixed point, we rewrite the flow equation (\ref{RG_Delta}) in terms of renormalized dimensionless quantities, which are defined by
\begin{equation}
\tilde{\psi}=k^{\zeta} \psi,
\end{equation}
\begin{equation}
\tilde{\Delta}_{2}^{\alpha \beta}(\tilde{\bvec{\psi}}) = \frac{16}{D} A_D k^{D-4+2\zeta} \Delta_{2}^{\alpha \beta}(\bvec{\psi}),
\end{equation}
\begin{equation}
\tilde{\Delta}_{3}^{\alpha \beta}(\tilde{\bvec{\psi}}) = \left(\frac{16}{D}\right)^2 {A_D}^2 k^{2D-6+2\zeta} \Delta_{3}^{\alpha \beta}(\bvec{\psi}),
\end{equation}
where $\zeta$ is the roughness exponent.
The flow equation of $\tilde{\Delta}^{\mu \nu}$ at $D=4-\epsilon$ is given by
\begin{eqnarray}
\partial_l \tilde{\Delta}^{\mu \nu}(\tilde{\bvec{\psi}}) &=& (\epsilon-2\zeta) \tilde{\Delta}^{\mu \nu}(\tilde{\bvec{\psi}}) + \zeta \tilde{\psi}^{\alpha} \partial_{\alpha} \tilde{\Delta}^{\mu \nu}(\tilde{\bvec{\psi}}) \nonumber \\
&&+ \partial_{\alpha} \partial_{\beta} \tilde{\Delta}^{\mu \nu}(\tilde{\bvec{\psi}}) (\tilde{\Delta}^{\beta \alpha}(\bvec{0})-\tilde{\Delta}^{\beta \alpha}(\tilde{\bvec{\psi}})) \nonumber \\
&&- \partial_{\alpha} \tilde{\Delta}^{\mu \beta}(\tilde{\bvec{\psi}}) \partial_{\beta} \tilde{\Delta}^{\nu \alpha}(\tilde{\bvec{\psi}}) -\tilde{\Delta}_3^{\mu \nu}(\tilde{\bvec{\psi}}).
\label{RG_Delta_eq}
\end{eqnarray}
Especially, for $N=1$, we have
\begin{eqnarray}
\partial_l \tilde{\Delta}(\tilde{\psi}) &=& (\epsilon-2\zeta) \tilde{\Delta}(\tilde{\psi}) + \zeta \tilde{\psi} \tilde{\Delta}'(\tilde{\psi}) + \tilde{\Delta}''(\tilde{\psi}) (\tilde{\Delta}(0)-\tilde{\Delta}(\tilde{\psi})) \nonumber \\
&&- \tilde{\Delta}'(\tilde{\psi})^2 -\tilde{\Delta}_3(\tilde{\psi}).
\label{RG_Delta_eq_N=1}
\end{eqnarray}
If we ignore the third cumulant $\tilde{\Delta}_3 = \mathcal{O}(\tilde{\Delta}^3)$, Eqs.~(\ref{RG_Delta_eq}) and (\ref{RG_Delta_eq_N=1}) are identical to those obtained in Refs.~\cite{Fisher-86} and \cite{Balents-93} by using the one-loop perturbative calculation.
To simplify the notation, we omit the tilde in $\tilde{\Delta}$ and $\tilde{\psi}$.

The second cumulant $\Delta^{\mu \nu}(\bvec{\psi})$ can be written as
\begin{eqnarray}
\Delta^{\mu \nu}(\bvec{\psi}) = \Delta_0(\rho) \delta^{\mu \nu} + \Delta_1(\rho) \psi^{\mu} \psi^{\nu},
\label{Delta0_Delta1}
\end{eqnarray}
where $\rho=|\bvec{\psi}|^2/2$.
By substituting Eq.~(\ref{Delta0_Delta1}) into Eq.~(\ref{RG_Delta_eq}), we have
\begin{eqnarray}
\Delta_1(\rho) = \Delta_0'(\rho),
\label{Delta1_Delta'0_eq}
\end{eqnarray}
which is obvious from the fact that $\Delta^{\mu \nu}(\bvec{\psi})$ can be expressed as the second derivative of the random potential cumulant (see Eq.~(\ref{Delta_R})).
If we ignore the third cumulant $\Delta_3$, the flow equation for $\Delta_0(\rho)$ is given by
\begin{eqnarray}
\partial_l \Delta_0(\rho) &=& (\epsilon-2\zeta) \Delta_0(\rho) + 2\zeta \rho \Delta_0'(\rho) \nonumber \\
&&+ [ (N+2) \Delta_0'(\rho) + 2 \rho \Delta_0''(\rho) ] ( \Delta_0(0)-\Delta_0(\rho) ) \nonumber \\
&&- 6 \rho \Delta_0'(\rho)^2 - 4 \rho^2 \Delta_0'(\rho) \Delta_0''(\rho).
\label{RG_Delta0_eq}
\end{eqnarray}

\subsubsection{Nonequilibrium case}

We next consider the nonequilibrium case ($v \neq 0$).
Below, the momentum cutoff $\Lambda$ is set to unity.
Considering the anisotropy of the system due to the driving, the transverse momentum $\mathbf{q}_{\perp}$ and longitudinal momentum $q_x$ are measured in units of $k$ and $k^2$,  respectively,
\begin{equation}
y_{\perp} = \frac{|\mathbf{q}_{\perp}|^2}{k^2}, \:\:\: y_{\parallel} = \frac{q_x^2}{k^4}.
\end{equation}
We employ an infrared cutoff function independent of $q_x$, 
\begin{equation}
R_k(\mathbf{q})=k^2 r (y_{\perp}) = k^2 (1-y_{\perp}) \Theta(1-y_{\perp}).
\label{cutoff_function_neq}
\end{equation}
The integrals in Eqs.~(\ref{def_L-}) and (\ref{def_L+}) are then calculated as follows:
\begin{eqnarray}
L_n^{\pm} = \frac{4}{D-1}A_{D-1} k^{D-2n+1} v^{-1} l_n^{\pm}(z_{k}),
\end{eqnarray}
where
\begin{equation}
z_{k} = v^{-2} k^2 = v^{-2} e^{-2l},
\label{def_z}
\end{equation}
which is related to the ratio of the longitudinal elastic term $\partial_x^2 \psi$ to the advection term $v \partial_x \psi$, and
\begin{eqnarray}
l_n^{-}(z) &=& \frac{n}{\pi} \int_{-\infty}^{\infty} dx \: (1+z x^2 + i x)^{-(n+1)}, \nonumber \\
l_n^{+}(z) &=& \frac{1}{\pi} \int_{-\infty}^{\infty} dx \: \sum_{j=1}^n (1+z x^2 + i x)^{-(n+1-j)} (1+z x^2 - i x)^{-j}.
\end{eqnarray}
One can easily check that $l_n^{+}(0)=1$ while $l_n^{-}(z) \sim z^n $ for a small $z$.

The renormalized dimensionless quantities are defined by
\begin{equation}
\tilde{\psi}=k^{\zeta_{\perp}} \psi,
\end{equation}
\begin{eqnarray}
\tilde{\Delta}_{2}^{\alpha \beta}(\tilde{\bvec{\psi}}) = \frac{4}{D-1}A_{D-1} v^{-1} k^{D-3+2\zeta_{\perp}} \Delta_{2}^{\alpha \beta}(\bvec{\psi}), 
\end{eqnarray}
\begin{eqnarray}
\tilde{\Delta}_{3}^{\alpha \beta}(\tilde{\bvec{\psi}}) = \left(\frac{4}{D-1}\right)^2 {A_{D-1}}^2 v^{-2} k^{2D-4+2\zeta_{\perp}} \Delta_{3}^{\alpha \beta}(\bvec{\psi}),
\end{eqnarray}
where $\zeta_{\perp}$ is the roughness exponent for the transverse directions.
The flow equation at $D=3-\epsilon$ is given by
\begin{eqnarray}
\partial_l \tilde{\Delta}^{\mu \nu}(\tilde{\bvec{\psi}}) &=& (\epsilon-2\zeta_{\perp}) \tilde{\Delta}^{\mu \nu}(\tilde{\bvec{\psi}}) + \zeta_{\perp} \tilde{\psi}^{\alpha} \partial_{\alpha} \tilde{\Delta}^{\mu \nu}(\tilde{\bvec{\psi}}) \nonumber \\
&&+ \partial_{\alpha} \partial_{\beta} \tilde{\Delta}^{\mu \nu}(\tilde{\bvec{\psi}}) (\tilde{\Delta}^{\beta \alpha}(\bvec{0})-\tilde{\Delta}^{\beta \alpha}(\tilde{\bvec{\psi}})) l_2^{+}(z_l) \nonumber \\
&&- \partial_{\alpha} \tilde{\Delta}^{\mu \beta}(\tilde{\bvec{\psi}}) \partial_{\beta} \tilde{\Delta}^{\nu \alpha}(\tilde{\bvec{\psi}}) l_2^{-}(z_l) -2\tilde{\Delta}_3^{\mu \nu}(\tilde{\bvec{\psi}}) l_1^{-}(z_l),
\label{RG_Delta_neq}
\end{eqnarray}
where $z_{l}$ is defined by Eq.~(\ref{def_z}).
Especially, for $N=1$, we have
\begin{eqnarray}
\partial_l \tilde{\Delta}(\tilde{\psi}) &=& (\epsilon-2\zeta_{\perp}) \tilde{\Delta}(\tilde{\psi}) + \zeta_{\perp} \tilde{\psi} \tilde{\Delta}'(\tilde{\psi}) + \tilde{\Delta}''(\tilde{\psi}) (\tilde{\Delta}(0)-\tilde{\Delta}(\tilde{\psi})) l_2^{+}(z_l) \nonumber \\
&&- \tilde{\Delta}'(\tilde{\psi})^2 l_2^{-}(z_l) - 2 \tilde{\Delta}_3(\tilde{\psi}) l_1^{-}(z_l).
\end{eqnarray}

Eq.~(\ref{RG_Delta_neq}) is not closed due to the presence of the third cumulant $\tilde{\Delta}_3 = \mathcal{O}(\tilde{\Delta}^3)$.
However, at weak disorder, this higher-order contribution is irrelevant to the fixed point because the coefficient $l_1^{-}(z_l)$ vanishes as $\sim e^{-2l}$ in the limit $l \to \infty$.
Therefore, remarkably, the flow equation (\ref{RG_Delta_neq}) is closed in the large length scale.
This conclusion does not rely on the specific form of the cutoff function $r (y_{\perp})$ in Eq.~(\ref{cutoff_function_neq}).
It is worth to note that, for the equilibrium case, since the coefficient of $\tilde{\Delta}_3$ is just a constant in the scaled form (see Eq.~(\ref{RG_Delta_eq})), the contributions of the higher-order cumulants can affect the value of $\zeta$.

Since we are interested in the fixed point, $z_l \sim e^{-2l}$ is set to zero from the beginning.
In the following, we omit the tilde in $\tilde{\Delta}$ and $\tilde{\psi}$ for simplicity.
By noting Eq.~(\ref{Delta0_Delta1}), the flow equations for $\Delta_0(\rho)$ and $\Delta_1(\rho)$ are given by
\begin{eqnarray}
\partial_l \Delta_0(\rho) &=& (\epsilon-2\zeta_{\perp}) \Delta_0(\rho) + 2\zeta_{\perp} \rho \Delta_0'(\rho) \nonumber \\
&&+ \bigl( N \Delta_0'(\rho) + 2 \Delta_1(\rho) + 2 \rho \Delta_0''(\rho) \bigr) \bigl( \Delta_0(0)-\Delta_0(\rho) \bigr) \nonumber \\
&&- 2 \rho \bigl( \Delta_0'(\rho) + 2 \rho \Delta_0''(\rho) \bigr) \Delta_1(\rho),
\label{RG_Delta0}
\end{eqnarray}
\begin{eqnarray}
\partial_l \Delta_1(\rho) &=& \epsilon \Delta_1(\rho) + 2\zeta_{\perp} \rho \Delta_1'(\rho) \nonumber \\
&&+ \bigl[ (N+4) \Delta_1'(\rho) + 2 \rho \Delta_1''(\rho) \bigr] \bigl( \Delta_0(0)-\Delta_0(\rho) \bigr) \nonumber \\
&&- 2 \Delta_1(\rho) \bigl( \Delta_1(\rho) + 5 \rho \Delta_1'(\rho) + 2 \rho^2 \Delta_1''(\rho) \bigr).
\label{RG_Delta1}
\end{eqnarray}
Note that Eq.~(\ref{Delta1_Delta'0_eq}) is no longer satisfied because a nonpotential random force is generated due to the breakdown of the detailed balance condition.

\section{Results}
\label{sec:Results}

First, we briefly review the results of the equilibrium case.
If we assume that $\Delta_0(\rho)$ is analytic near $\rho=0$, we obtain the flow equations of $\Delta_0(0)$ and $\Delta_0'(0)$ from Eq.~(\ref{RG_Delta0_eq}),
\begin{eqnarray}
\partial_l \Delta_0(0) &=& (\epsilon-2\zeta) \Delta_0(0), \nonumber \\
\partial_l \Delta_0'(0) &=& \epsilon \Delta_0'(0) - (N+8) \Delta_0'(0)^2,
\label{RG_Delta(0)_Delta'(0)_eq}
\end{eqnarray}
where $\epsilon=4-D$.
At a fixed point, the first equation yields $\zeta=\epsilon/2$, which is nothing but the dimensional reduction value.
On the other hand, the second equation does not have any fixed point solution because $\Delta_0'(0)$ should be negative.
This means that any analytic function cannot be a fixed point of Eq.~(\ref{RG_Delta0_eq}).
In fact, from the second equation of Eq.~(\ref{RG_Delta(0)_Delta'(0)_eq}), one can see that $\Delta_0'(0)$ diverges to minus infinity at a finite renormalization scale $l_{\mathrm{L}}$, which is known as the Larkin scale \cite{Fisher-86}.
For $l>l_{\mathrm{L}}$, $\Delta_0(\rho)$ develops a square root cusp at the origin, $\Delta_0(\rho) - \Delta_0(0) \sim -\rho^{1/2} $.
Then, the first equation of Eq.~(\ref{RG_Delta(0)_Delta'(0)_eq}) is modified as
\begin{eqnarray}
\partial_l \Delta_0(0) = (\epsilon-2\zeta) \Delta_0(0) - \lim_{\rho \to 0} [(N+8) \rho \Delta_0'(\rho)^2 + 6 \rho^2 \Delta_0'(\rho) \Delta_0''(\rho)].
\end{eqnarray}
Since $\lim_{\rho \to 0} [...]$ is positive, we have $\zeta < \zeta_{\mathrm{DR}} = \epsilon/2$.
Thus, the strength of the cusp in the renormalized disorder cumulant is directly related to the amount of breakdown of the dimensional reduction.

We next consider the nonequilibrium case, where the flow equations are given by Eqs.~(\ref{RG_Delta0}) and (\ref{RG_Delta1}) with $\epsilon=3-D$.
First, note that, if the derivative of $\Delta_0(\rho)$ is finite at $\rho=0$, we have $\zeta_{\perp}=\epsilon/2$ by setting $\rho=0$ in Eq.~(\ref{RG_Delta0}), which is the dimensional reduction value.
However, this assumption is incorrect.
In fact, if we assume that the fixed point functions $\Delta_0(\rho)$ and $\Delta_1(\rho)$ are analytic at $\rho=0$, we have,
\begin{eqnarray}
0 &=& \epsilon \Delta_0'(0) - N \Delta_0'(0)^2 -4 \Delta_0'(0) \Delta_1(0), \nonumber \\
0 &=& \epsilon \Delta_1(0) - 2 \Delta_1(0)^2,
\end{eqnarray}
where the first equation has been obtained from the first derivative of Eq.~(\ref{RG_Delta0}).
This set of equations has no solution satisfying $\Delta_0'(0), \Delta_1(0) < 0$.
Therefore, we seek for a nonanalytic fixed point which behaves as $\Delta_0(\rho) - \Delta_0(0) \sim -\rho^{1/2} $ near $\rho=0$.

We expand the nonanalytic solution as
\begin{eqnarray}
\Delta_0(\rho) &=& a_0 + a_1 \rho^{1/2} + a_2 \rho +..., \nonumber \\
\Delta_1(\rho) &=& b_{-1} \rho^{-1/2} +  b_0 + b_1 \rho^{1/2} + b_2 \rho +....
\label{Delta_expansion}
\end{eqnarray}
For the equilibrium case, we have a relation $b_n=(n+2)a_{n+2}/2$ from Eq.~(\ref{Delta1_Delta'0_eq}).
For the nonequilibrium case, by substituting Eq.~(\ref{Delta_expansion}) into Eqs.~(\ref{RG_Delta0}) and (\ref{RG_Delta1}), we find that
\begin{equation}
b_{-1} = 0,
\end{equation}
because the right-hand side of Eq.~(\ref{RG_Delta1}) yields $(1/2)(N+1)a_1 b_{-1} \rho^{-1}$.
This means that, at the fixed point, $\Delta_1(0)$ is finite in contrast to the equilibrium case where it diverges as $\rho^{-1/2}$.

\begin{figure}
 \centering
 \includegraphics[width=0.7\textwidth]{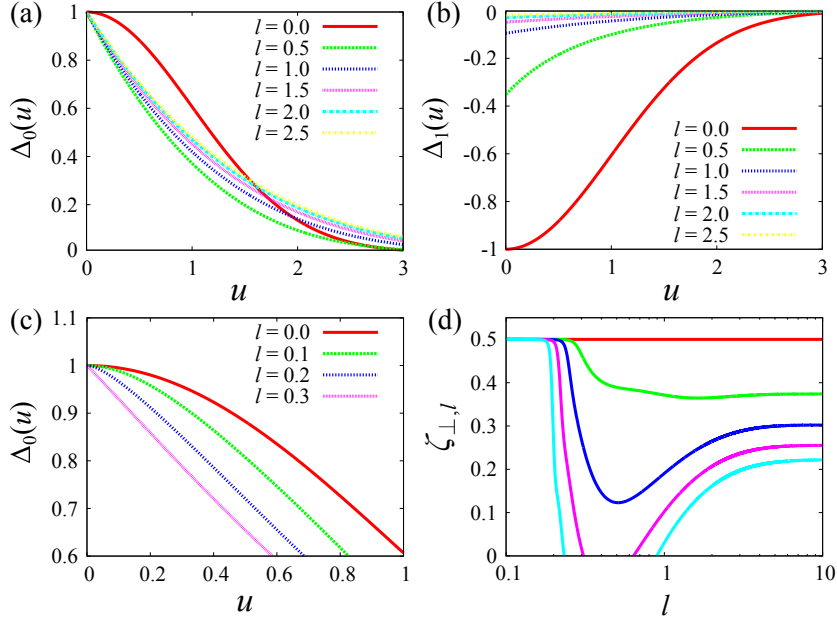}
 \caption{(a), (b): RG evolution of $\Delta_0(u)$ and $\Delta_1(u)$ for $N=2$.
 The initial condition is as follows: $\Delta_{0,l=0}(u)=-\Delta_{1,l=0}(u)=\exp(-u^2/2)$.
 The values of the renormalization scale are $l=0.0,0.5,1.0,1.5,2.0,2.5$.
 (c): RG evolution of $\Delta_0(u)$ in the small-$u$ region.
 The values of the renormalization scale are $l=0.0,0.1,0.2,0.3$.
 (d): Scale-dependent roughness exponent $\zeta_{\perp,l}$ in a semi-log plot.
 From the top to the bottom, $N=1.0,1.5,2.0,2.5,3.0$.
 In all calculations, $\epsilon=3-D$ is set to unity.}
 \label{fig-1}
\end{figure}

In numerical integration of these equations, it is convenient to introduce a parameter $u$ by $\rho=u^2/2$.
Then, Eqs.~(\ref{RG_Delta0}) and (\ref{RG_Delta1}) are rewritten as
\begin{eqnarray}
\partial_l \Delta_0(u) &=& (\epsilon-2\zeta_{\perp}) \Delta_0(u) + \zeta_{\perp} u \Delta_0'(u) \nonumber \\
&&+ \bigl[ (N-1) u^{-1} \Delta_0'(u) + 2 \Delta_1(u) + \Delta_0''(u) \bigr] \bigl( \Delta_0(0)-\Delta_0(u) \bigr) \nonumber \\
&&- u^2 \Delta_0''(u) \Delta_1(u),
\label{RG_Delta0_u}
\end{eqnarray}
\begin{eqnarray}
\partial_l \Delta_1(u) &=& \epsilon \Delta_1(u) + \zeta_{\perp} u \Delta_1'(u) \nonumber \\
&&+ \bigl[ (N+3) u^{-1} \Delta_1'(u) + \Delta_1''(u) \bigr] \bigl( \Delta_0(0)-\Delta_0(u) \bigr) \nonumber \\
&&- ( 2 \Delta_1(u) + 4 u \Delta_1'(u) + u^2 \Delta_1''(u)) \Delta_1(u).
\label{RG_Delta1_u}
\end{eqnarray}
We seek for a nonanalytic fixed point which has a linear cusp at the origin; $\Delta_0(u)=a_0+(a_1/\sqrt{2})|u|$, $\Delta_1(u)=b_0+(b_1/\sqrt{2})|u|$.
The roughness exponent $\zeta_{\perp}$ should be chosen such that $\Delta_0(u)$ and $\Delta_1(u)$ attain a stable fixed point in the limit $l \to \infty$.
Therefore, we define the scale-dependent roughness exponent $\zeta_{\perp,l}$ from the condition that $\Delta_0(0)$ is constant along the RG flow.
From Eq.~(\ref{RG_Delta0_u}), we have
\begin{equation}
\zeta_{\perp,l} = \frac{1}{2} \bigl[ \epsilon-(N-1)\Delta_0'(0^+)^2 \bigr],
\label{Scale_dependent_zeta}
\end{equation}
where $\Delta_0(0)$ is set to unity.
We first consider the short-range correlated disorder Eq.~(\ref{R_short_range}), thus we assume that $\Delta_0(u)$ and $\Delta_1(u)$ decay as $\sim \exp(-u^2/2)$ for large $u$.

We numerically integrate Eqs.~(\ref{RG_Delta0_u}) and (\ref{RG_Delta1_u}).
We set the space and time discretization as $\Delta u=5\times 10^{-3}$ and $\Delta l=10^{-5}$, respectively.
The upper panels (a) and (b) in Fig.~\ref{fig-1} show the RG evolution of $\Delta_0(u)$ and $\Delta_1(u)$ for $N=2$.
We have employed the following initial condition: $\Delta_{0,l=0}(u)=-\Delta_{1,l=0}(u)=\exp(-u^2/2)$.
From the panel (c) in Fig.~\ref{fig-1}, one can clearly see that $\Delta_0(u)$ develops a linear cusp at a finite renormalization scale $l_{\mathrm{L}}$.
In the limit $l \to \infty$, $\Delta_0(u)$ converges to a non-zero fixed point $\Delta_0^*(u)$, in contrast, $\Delta_1(u)$ flows to zero.
Note that the fixed function $\Delta_0^*(u)$ is independent of the initial condition.
The panel (d) in Fig.~\ref{fig-1} shows the scale-dependent roughness exponent $\zeta_{\perp,l}$ for $N=1.0,1.5,2.0,2.5,3.0$ from the top to the bottom.
For $N=1$, $\zeta_{\perp,l}=\epsilon/2$ for all $l$, which also follows from Eq.~(\ref{Scale_dependent_zeta}).
For $N>1$, $\zeta_{\perp,l}$ has the dimensional reduction value $\epsilon/2$ for $0 \leq l \leq l_{\mathrm{L}}$, but it deviates from $\epsilon/2$ for $l_{\mathrm{L}} \leq l$, due to the generation of the cusp in $\Delta(u)$.

\begin{figure}
 \centering
 \includegraphics[width=0.5\textwidth]{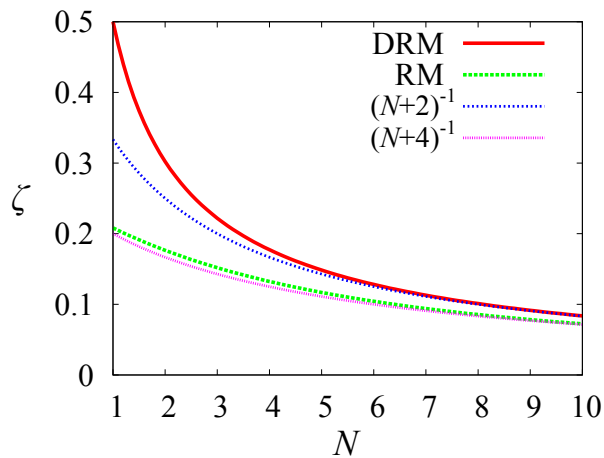}
 \caption{Roughness exponents $\zeta$ for the short-range correlated disorder.
 The red solid line represents {\it exact} $\zeta_{\perp}$ of the DRM at $D=3-\epsilon$.
 The green dashed line represents {\it one-loop} $\zeta$ of the RM at $D=4-\epsilon$.
 The thin dotted lines display $(N+2)^{-1}$ and $(N+4)^{-1}$.
 $\epsilon$ is set to unity.}
 \label{fig-2}
\end{figure}

\begin{figure}
 \centering
 \includegraphics[width=0.5\textwidth]{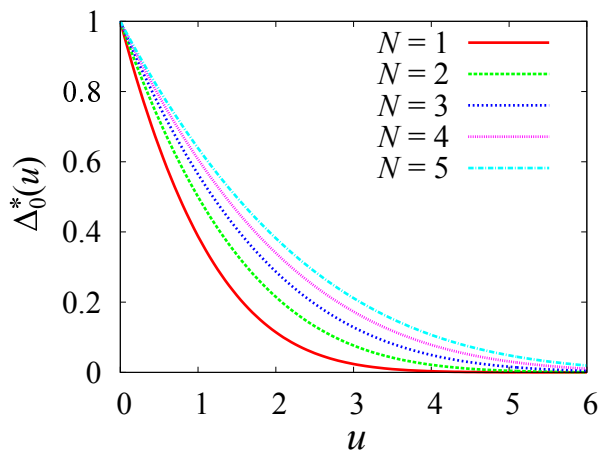}
 \caption{Fixed function $\Delta_0^*(u)$ for the short-range correlated disorder.
 The numbers of the field components are $N=1,2,3,4,5$ from the bottom to the top.}
 \label{fig-3}
\end{figure}

We next calculate the fixed point value of the roughness exponent $\zeta_{\perp} = \lim_{l \to \infty} \zeta_{\perp,l}$.
Since $\Delta_1(u)$ goes to zero in the limit $l \to \infty$, the fixed function $\Delta_0^*(u)$ satisfies the following equation: 
\begin{eqnarray}
0 &=& (\epsilon-2\zeta_{\perp}) \Delta_0(u) + \zeta_{\perp} u \Delta_0'(u) \nonumber \\
&&+ \bigl[ (N-1) u^{-1} \Delta_0'(u) + \Delta_0''(u) \bigr] \bigl( \Delta_0(0)-\Delta_0(u) \bigr).
\label{RG_Delta0_FP_u}
\end{eqnarray}
Note that if $\Delta_0(u)$ is a fixed point, then $\kappa^2 \Delta_0(u/\kappa)$ is also a fixed point for any $\kappa>0$.
We use this property to set $\Delta_0(0)=1$.
For a fixed $\zeta_{\perp}$, Eq.~(\ref{RG_Delta0_FP_u}) can be numerically integrated from $u=0$ with the initial condition $\Delta_0(0)=1$ and $\Delta_0'(0^+)=\sqrt{(\epsilon-2\zeta_{\perp})/(N-1)}$, where the second condition comes from Eq.~(\ref{Scale_dependent_zeta}).
We seek for $\zeta_{\perp}$ which leads to an exponentially decaying $\Delta_0(u)$.
For $N=1$, $\zeta_{\perp}=\epsilon/2$ and $\Delta_0'(0^+)$ is tuned so as to obtain the short-range fixed function.
Figure \ref{fig-2} shows the roughness exponents $\zeta$ as functions of $N$.
The red solid line represents $\zeta_{\perp}$ of the DRM at $D=3-\epsilon$.
The green dashed line represents one-loop $\zeta$ of the RM at $D=4-\epsilon$, which is calculated from Eq.~(\ref{RG_Delta0_eq}).
Note that, for the RM, $\zeta$ is correct within the first order of $\epsilon$.
The contribution from the third-order cumulant $\Delta_3^{\mu \nu}$ in Eq.~(\ref{RG_Delta_eq}) yields the correction of $\mathcal{O}(\epsilon^2)$ to $\zeta$.
In contrast, for the DRM, $\zeta_{\perp}$ obtained from Eq.~(\ref{RG_Delta0_FP_u}) is expected to be exact for $1<D<3$ because the third-order cumulant $\Delta_3^{\mu \nu}$ in Eq.~(\ref{RG_Delta_neq}) vanishes in the large-scale limit due to the exponential factor $l_1^{-}(z_l) \sim e^{-2l}$.
For the RM with $N=1$, we have $\zeta=0.208\epsilon$, which is the well-known value \cite{Fisher-86}.
For the DRM with $N=1$, we have $\zeta_{\perp}=\epsilon/2$ as mentioned above.
Figure \ref{fig-3} shows the fixed function $\Delta_0^*(u)$ for different values of $N$. 
Note that, for $N=1$, although $\Delta_0^*(u)$ has a cusp, the roughness exponent is equal to its dimensional reduction value  .

Let us consider the asymptotic behavior of $\zeta_{\perp}$ for large $N$.
The fixed point equation for $\Delta_0(\rho)$ is given by
\begin{eqnarray}
0 &=& (\epsilon-2\zeta_{\perp}) \Delta_0(\rho) + 2\zeta_{\perp} \rho \Delta_0'(\rho) \nonumber \\
&&+ \bigl( N \Delta_0'(\rho) + 2 \rho \Delta_0''(\rho) \bigr) \bigl( \Delta_0(0)-\Delta_0(\rho) \bigr).
\label{RG_Delta0_FP}
\end{eqnarray}
By introducing $\delta(\rho)=(\epsilon-2\zeta_{\perp})^{-1} N \Delta_0(\rho)$, Eq.~(\ref{RG_Delta0_FP}) can be rewritten as
\begin{eqnarray}
0 = \delta(\rho) + \beta \rho \delta'(\rho) + \bigl( \delta'(\rho) + \mu \rho \delta''(\rho) \bigr) \bigl( \delta(0)-\delta(\rho) \bigr),
\label{RG_delta_FP}
\end{eqnarray}
where $\beta=2\zeta_{\perp}/(\epsilon-2\zeta_{\perp})$ and $\mu=2/N$.
For $\rho \gg 1$, the quadratic term $\mathcal{O}(\delta^2)$ can be neglected because $\delta(\rho)$ decays exponentially for large $\rho$,
\begin{eqnarray}
0 = \delta(\rho) + \beta \rho \delta'(\rho) + \bigl( \delta'(\rho) + \mu \rho \delta''(\rho) \bigr) \delta(0).
\label{RG_delta_FP_linear}
\end{eqnarray}
Below, we set $\delta(0)=1$.
Since $\beta, \mu = \mathcal{O}(N^{-1})$, in the regime $1 \ll \rho \ll N$, Eq.~(\ref{RG_delta_FP_linear}) can be reduced to
\begin{equation}
0 = \delta(\rho) + \delta'(\rho),
\end{equation}
thus we have 
\begin{equation}
\delta(\rho) \sim e^{-\rho}.
\label{FP_solution_1}
\end{equation}
In the regime $N \ll \rho$, Eq.~(\ref{RG_delta_FP_linear}) can be reduced to
\begin{equation}
0 = \beta \rho \delta'(\rho) + \mu \rho \delta''(\rho),
\end{equation}
thus we have
\begin{equation}
\delta(\rho) \sim e^{-(\beta/\mu)\rho}.
\label{FP_solution_2}
\end{equation}
These two solutions (\ref{FP_solution_1}) and (\ref{FP_solution_2}) should match around $\rho \sim N$.
Therefore, we have $\beta = \mu$, which implies the following asymptotic behavior:
\begin{equation}
\zeta_{\perp} = \frac{\epsilon}{N+2}.
\end{equation}
For the equilibrium case Eq.~(\ref{RG_Delta0_eq}), the similar argument leads to $\zeta = \epsilon/(N+4)$ \cite{Balents-93}.
These two asymptotic functions are displayed in Fig.~\ref{fig-2} by the thin dotted lines.

We next consider the long-range correlated disorder defined by Eq.~(\ref{R_long_range}).
We seek for the fixed function which exhibits a power-law decay, $\Delta_0(\rho) \sim \rho^{-\gamma}$.
For $\rho \gg 1$, since $\Delta'(\rho) \sim \rho \Delta''(\rho) \ll \Delta(\rho) \sim \rho \Delta'(\rho)$, Eq.~(\ref{RG_Delta0_FP}) can be reduced to
\begin{equation}
0 = (\epsilon-2\zeta_{\perp}) \Delta_0(\rho) + 2\zeta_{\perp} \rho \Delta_0'(\rho).
\end{equation}
Therefore, the roughness exponent is given by
\begin{equation}
\zeta_{\perp}^{\mathrm{LR}}= \frac{\epsilon}{2(1+\gamma)},
\end{equation}
which is the same as that of the RM \cite{Balents-93}.
Especially, for the random field case $\gamma=1/2$, we have $\zeta_{\perp}^{\mathrm{LR}}=\epsilon/3$.
This result is expected to be exact for $1<D<3$.

We define $\gamma_{\mathrm{c}}$ by $\zeta_{\perp}^{\mathrm{LR}}=\zeta_{\perp}^{\mathrm{SR}}$, where $\zeta_{\perp}^{\mathrm{SR}}$ is the roughness exponent for the short-range case.
For $\gamma>\gamma_{\mathrm{c}}$, there is no fixed function which decays as $\Delta_0(\rho) \sim \rho^{-\gamma}$.
Therefore, in this case, we expect that the long-range cumulant flows to the short-range fixed point shown in Fig.~\ref{fig-3}.
Especially, when $N=1$, $\zeta_{\perp}=\epsilon/2$ for any $\gamma>0$.

\section{Concluding remarks}
\label{sec:Conclusions}

In this study, we discussed the dimensional reduction which relates the large-scale behavior of the driven disordered systems in $D$ dimensions to that of the pure systems in $D-1$ dimensions.
However, it can break down due to the presence of multiple stationary states.
We also performed the FRG analysis of the $N$-component elastic manifold transversely driven in a random potential and found that the cusp in the renormalized disorder cumulant causes the failure of the dimensional reduction.
Remarkably, for $N=1$, the roughness exponent is the same as its dimensional reduction value despite of the presence of the cusp in the disorder cumulant.

We provide a brief explanation for the physical mechanism of the generation of the cusp in the renormalized disorder cumulant.
In equilibrium, when a force is exerted on the elastic manifold at zero temperature, it exhibits a discontinuous motion known as avalanche dynamics, if the force is strong enough to overcome the pinning potential.
The critical force required to depin the manifold is proportional to the first derivative of the random force cumulant $|\Delta'(\phi=0^+)|$.
The presence of the multiple stationary states, the generation of the cusp in the renormalized disorder cumulant, and the onset of the avalanche behavior are closely related each other, and they lead to the breakdown of the dimensional reduction.
In driven systems, the situation is quite similar to that in equilibrium.
Suppose that an additional force is exerted on the driven random manifold to the direction perpendicular to the driving velocity.
At zero temperature, a nonzero force is required to depin the manifold along the force direction and its value corresponds to the strength of the cusp in the renormalized disorder cumulant.
Recall that in equilibrium the presence of the avalanche behavior leads to the spontaneous breaking of the supersymmetry, on which the dimensional reduction is based \cite{Tissier-12-1,Tissier-12-2}.
In contrast, for the nonequilibrium case, one should keep in mind that there is no connection between the supersymmetry breaking and the onset of the avalanche behavior.

We remark about the large-$N$ limit of the DRM.
By rescaling $N \Delta_0(\rho) \to \Delta_0(\rho)$, Eq.~(\ref{RG_Delta0}) can be reduced to
\begin{eqnarray}
\partial_l \Delta_0(\rho) = (\epsilon-2\zeta_{\perp}) \Delta_0(\rho) + 2\zeta_{\perp} \rho \Delta_0'(\rho) + \Delta_0'(\rho) \bigl( \Delta_0(0)-\Delta_0(\rho) \bigr),
\label{RG_Delta0_large_N}
\end{eqnarray}
in the limit $N \to \infty$.
This FRG equation is the same as that of the RM in the large-$N$ limit at $D=4-\epsilon$ \cite{Balents-93}.
The roughness exponent is $\zeta_{\perp}=0$ for the short-range correlated disorder and $\zeta_{\perp}=\epsilon/3$ for the random field type disorder.
It is worth to note that the DRM can be directly solved in this limit and one can derive an exact self-consistent equation for the effective action as done in Ref.~\cite{LeDoussal-03} for the RM.
This self-consistent equation has the same structure as that of the RM except that the free propagator is replaced with the nonequilibrium one.

It is interesting to make comparison between the results for the DRM and those of the driven random field $O(N)$ model (DRF$O(N)$M), which is defined by Eqs.~(\ref{General_Hamiltonian}) and (\ref{General_EM}) with $U(\bvec{\phi})=\lambda(|\bvec{\phi}|^2-1)^2$ and $\Delta(\bvec{\phi})=\Delta_0$.
The FRG analysis of the DRF$O(N)$M was performed in Ref.~\cite{Haga-17}, where we calculated the critical exponent $\eta$ which characterizes the power-law decay of the correlation function at the critical point.
As a result, we found that the dimensional reduction recovers for sufficiently large field component number $N$.
This contrasts with the case of the DRM, where the nonperturbative effect responsible for the breakdown of the dimensional reduction is most pronounced in the large $N$ limit, $\zeta_{\perp} \to 0$.

Note that the DRF$O(2)$M can be mapped to the single component DRM with a periodic random potential by using a phase parameter $u$ defined by $\bvec{\phi}=(\phi^1,\phi^2)=(\cos u, \sin u)$, if topological defects (vortices) are ignored.
Since we have shown that the dimensional reduction holds for the single component DRM, the long-distance physics of the DRF$O(2)$M is expected to be identical to that of the pure $O(2)$ (XY) model in a reduced dimension.
It may be recalled that the two-dimensional XY model exhibits quasi-long-range order (QLRO) at low temperatures, wherein the correlation function shows power-law decay with a continuously varying exponent.
The transition from this QLRO phase to a disordered phase is known as the Kosterlitz-Thouless (KT) transition.
The dimensional reduction suggests that the DRF$O(2)$M exhibits QLRO and the KT transition in three dimensions.
The detailed investigation for this remarkable possibility is given in Refs.~\cite{Haga-15,Haga-18}.

In this paper, we restricted our attention to the case of zero temperature.
At finite temperatures, the thermal noise enables the manifold to escape a stationary state and the avalanche behavior is smeared.
In fact, the diffusion-like term appears in the flow equation for the disorder cumulant (see Eq.~(\ref{RG_Delta})) and it smooths out the cusp.
For disordered systems in equilibrium, since the temperature is irrelevant, the zero-temperature fixed point always controls the long-distance physics of the system \cite{Nattermann-97}.
For driven disordered systems, in contrast, the breakdown of the detailed balance condition can lead to a generation of a temperature and a fixed point with a finite temperature appears \cite{LeDoussal-98}.
This means that an infinitesimally small thermal noise can change the large-scale behavior of the system.
It is an interesting future problem to understand how the finite-temperature behavior is connected to the zero-temperature one when the temperature decreases to zero.
In Sec.~3, we have argued that, at zero temperature, the stationary state of the driven disordered system is formally identical to a dynamical solution of the lower dimensional pure system.
In this argument, the random force is mapped to an effective thermal noise with a temperature $\Delta(0)/(2v)$.
One should not confuse this fictitious temperature with the real temperature characterizing the intensity of the thermal noise $\xi$ in Eq.~(\ref{General_EM}).
Since the effective thermal noise resulting from the dimensional reduction argument plays no role in the real activation dynamics of the manifold, it does not lead to the rounding of the cusp in the disorder cumulant.

\ack
The author acknowledges Gilles Tarjus and Matthieu Tissier for useful discussions.
The present study was supported by JSPS KAKENHI Grant No. 15J01614, a Grant-in-Aid for JSPS Fellows.

\section*{Appendix}

In this Appendix, we present the detailed derivation of the flow equation for $\Delta_2$, Eq.~(\ref{RG_Delta2_dimensionfull-1}).
For simplicity, we employ the following notations for the functional derivatives:
\begin{eqnarray}
\Gamma_{2,\psi^{\alpha} \hat{\psi}^{\beta} \hat{\psi}^{\gamma}}^{(21)}[\Psi_1,\Psi_2] = \frac{\delta^3 \Gamma_2[\Psi_1,\Psi_2]}{\delta \psi_1^{\alpha} \delta \hat{\psi}_1^{\beta} \delta \hat{\psi}_2^{\gamma}}, \nonumber \\
\Gamma_{3,\hat{\psi}^{\alpha} \psi^{\beta} \hat{\psi}^{\mu} \hat{\psi}^{\nu}}^{(121)}[\Psi_1,\Psi_2,\Psi_3] = \frac{\delta^4 \Gamma_2[\Psi_1,\Psi_2,\Psi_3]}{\delta \hat{\psi}_1^{\alpha} \delta \psi_2^{\beta} \delta \hat{\psi}_2^{\mu} \delta \hat{\psi}_3^{\nu}},
\end{eqnarray}
where the subscript ``$k$'' is omitted.
From Eqs.~(\ref{Exact_flow_Gamma_2}) and (\ref{RGeq_Delta-0}), we have
\begin{eqnarray}
\partial_l \Delta_2^{\mu \nu}(\bvec{\psi}_a, \bvec{\psi}_b) &=& -\frac{1}{2} \int_{\mathbf{q},\omega} \partial_l R_k(\mathbf{q}) \Bigl[(\mathrm{A})+(\mathrm{B}-1)+(\mathrm{B}-2) \nonumber \\
&&+(\mathrm{B}-3)+(\mathrm{C})+\mathrm{perm}(\bvec{\psi}_a, \bvec{\psi}_b)\Bigr],
\label{Appendix-RG_Delta2_dimensionfull-1}
\end{eqnarray}
\begin{eqnarray}
(\mathrm{A}) &=& \mathrm{P}_{\hat{\psi}^{\alpha} \psi^{\alpha}} \Gamma_{2,\psi^{\alpha} \psi^{\alpha} \hat{\psi}^{\mu} \hat{\psi}^{\nu}}^{(31)}[\Psi_a,\Psi_b] \mathrm{P}_{\psi^{\alpha} \psi^{\alpha}} \nonumber \\
&&+ \mathrm{P}_{\psi^{\alpha} \psi^{\alpha}} \Gamma_{2,\psi^{\alpha} \psi^{\alpha} \hat{\psi}^{\mu} \hat{\psi}^{\nu}}^{(31)}[\Psi_a,\Psi_b] \mathrm{P}_{\psi^{\alpha} \hat{\psi}^{\alpha}}, \nonumber
\end{eqnarray}
\begin{eqnarray}
(\mathrm{B}-1) &=& 2 \mathrm{P}_{\hat{\psi}^{\alpha} \psi^{\alpha}} \Gamma_{2,\psi^{\alpha} \psi^{\beta} \hat{\psi}^{\mu} \hat{\psi}^{\nu}}^{(31)}[\Psi_a,\Psi_b] \mathrm{P}_{\psi^{\beta} \hat{\psi}^{\beta}} \Gamma_{2,\hat{\psi}^{\beta} \hat{\psi}^{\alpha}}^{(11)}[\Psi_a,\Psi_a] \mathrm{P}_{\hat{\psi}^{\alpha} \psi^{\alpha}}, \nonumber
\end{eqnarray}
\begin{eqnarray}
(\mathrm{B}-2) &=& 2 \mathrm{P}_{\hat{\psi}^{\alpha} \psi^{\alpha}} \Gamma_{2,\psi^{\alpha} \hat{\psi}^{\mu} \psi^{\beta} \hat{\psi}^{\nu}}^{(22)}[\Psi_a,\Psi_b] \mathrm{P}_{\psi^{\beta} \hat{\psi}^{\beta}} \Gamma_{2,\hat{\psi}^{\beta} \hat{\psi}^{\alpha}}^{(11)}[\Psi_b,\Psi_a] \mathrm{P}_{\hat{\psi}^{\alpha} \psi^{\alpha}}, \nonumber
\end{eqnarray}
\begin{eqnarray}
(\mathrm{B}-3) &=& 2 \mathrm{P}_{\hat{\psi}^{\alpha} \psi^{\alpha}} \Gamma_{2,\psi^{\alpha} \hat{\psi}^{\mu} \hat{\psi}^{\beta}}^{(21)}[\Psi_a,\Psi_b] \mathrm{P}_{\hat{\psi}^{\beta} \psi^{\beta}} \Gamma_{2,\psi^{\beta} \hat{\psi}^{\nu} \hat{\psi}^{\alpha}}^{(21)}[\Psi_b,\Psi_a] \mathrm{P}_{\hat{\psi}^{\alpha} \psi^{\alpha}}, \nonumber
\end{eqnarray}
\begin{eqnarray}
(\mathrm{C}) &=&  \mathrm{P}_{\hat{\psi}^{\alpha} \psi^{\alpha}} \Gamma_{3,\psi^{\alpha} \hat{\psi}^{\mu} \hat{\psi}^{\alpha} \hat{\psi}^{\nu}}^{(211)}[\Psi_a,\Psi_a,\Psi_b] \mathrm{P}_{\hat{\psi}^{\alpha} \psi^{\alpha}} \nonumber \\
&&+ \mathrm{P}_{\psi^{\alpha} \hat{\psi}^{\alpha}} \Gamma_{3,\hat{\psi}^{\alpha} \psi^{\alpha} \hat{\psi}^{\mu} \hat{\psi}^{\nu}}^{(121)}[\Psi_a,\Psi_a,\Psi_b] \mathrm{P}_{\psi^{\alpha} \hat{\psi}^{\alpha}}, \nonumber
\end{eqnarray}
where the summation over repeated indices $\alpha, \beta$ is assumed.
In Eq.~(\ref{Appendix-RG_Delta2_dimensionfull-1}), we have omitted the terms which vanish when they are evaluated at a uniform field: $ \bvec{\psi}_{1,rt} \equiv \bvec{\psi}_1,...,\bvec{\psi}_{p,rt} \equiv \bvec{\psi}_p $ and  $ \bvec{\hat{\psi}}_{1,rt} \equiv 0,.., \bvec{\hat{\psi}}_{p,rt} \equiv 0 $.
For example, the term containing $(\Gamma_{2}^{(21)}[\Psi_1,\Psi_1]+\Gamma_{2}^{(12)}[\Psi_1,\Psi_1])$, which comes from the derivative of the second term in Eq.~(\ref{Exact_flow_Gamma_2}), vanishes.
The one-replica propagator $\mathrm{P}$ is written as
\begin{eqnarray}
\mathrm{P}_{\hat{\psi}^{\alpha} \psi^{\beta}}(q,q') = P_{21}(\mathbf{q},\omega) \delta^{\alpha \beta} \delta(\mathbf{q}+\mathbf{q}') \delta(\omega+\omega'),
\end{eqnarray}
and so on, where $P_{ij}(\mathbf{q},\omega)$ is given by Eq.~(\ref{propagator_component}) and we have omitted the trivial factors $2\pi$ associated with the delta function.

By using Eq.~(\ref{Gamma_n}), each functional derivative in Eq.~(\ref{Appendix-RG_Delta2_dimensionfull-1}) is calculated as follows:
\begin{eqnarray}
\Gamma_{2,\hat{\psi}^{\alpha}(q_1) \hat{\psi}^{\beta}(q_2)}^{(11)}[\Psi_1,\Psi_2] = \Delta_2^{\alpha \beta}(\bvec{\psi}_1,\bvec{\psi}_2) \delta(\mathbf{q}_1+\mathbf{q}_2) \delta(\omega_1) \delta(\omega_2),
\label{Appendix_Gamma_derivative-1}
\end{eqnarray}
\begin{eqnarray}
&&\Gamma_{2,\psi^{\alpha}(q_1) \hat{\psi}^{\mu}(q_2) \hat{\psi}^{\beta}(q_3)}^{(21)}[\Psi_1,\Psi_2] \nonumber \\ 
&&= \partial_{\psi_1^{\alpha}} \Delta_2^{\mu \beta}(\bvec{\psi}_1,\bvec{\psi}_2) \delta(\mathbf{q}_1+\mathbf{q}_2+\mathbf{q}_3) \delta(\omega_1+\omega_2) \delta(\omega_3),
\label{Appendix_Gamma_derivative-2}
\end{eqnarray}
\begin{eqnarray}
&&\Gamma_{2,\psi^{\alpha}(q_1) \psi^{\beta}(q_2) \hat{\psi}^{\mu}(q_3) \hat{\psi}^{\nu}(q_4)}^{(31)}[\Psi_1,\Psi_2] \nonumber \\
&&= \partial_{\psi_1^{\alpha}} \partial_{\psi_1^{\beta}} \Delta_2^{\mu \nu}(\bvec{\psi}_1,\bvec{\psi}_2) \delta(\mathbf{q}_1+\mathbf{q}_2+\mathbf{q}_3+\mathbf{q}_4) \delta(\omega_1+\omega_2+\omega_3) \delta(\omega_4),
\label{Appendix_Gamma_derivative-3}
\end{eqnarray}
\begin{eqnarray}
&&\Gamma_{2,\psi^{\alpha}(q_1) \hat{\psi}^{\mu}(q_2) \psi^{\beta}(q_3) \hat{\psi}^{\nu}(q_4)}^{(22)}[\Psi_1,\Psi_2] \nonumber \\
&&= \partial_{\psi_1^{\alpha}} \partial_{\psi_2^{\beta}} \Delta_2^{\mu \nu}(\bvec{\psi}_1,\bvec{\psi}_2) \delta(\mathbf{q}_1+\mathbf{q}_2+\mathbf{q}_3+\mathbf{q}_4) \delta(\omega_1+\omega_2) \delta(\omega_3+\omega_4),
\label{Appendix_Gamma_derivative-4}
\end{eqnarray}
\begin{eqnarray}
&&\Gamma_{3,\psi^{\alpha}(q_1) \hat{\psi}^{\mu}(q_2) \hat{\psi}^{\alpha}(q_3) \hat{\psi}^{\nu}(q_4)}^{(211)}[\Psi_1,\Psi_2,\Psi_3] \nonumber \\
&&= \partial_{\psi_1^{\alpha}} \Delta_3^{\mu \alpha \nu}(\bvec{\psi}_1,\bvec{\psi}_2,\bvec{\psi}_3) \delta(\mathbf{q}_1+\mathbf{q}_2+\mathbf{q}_3+\mathbf{q}_4) \delta(\omega_1+\omega_2) \delta(\omega_3) \delta(\omega_4).
\label{Appendix_Gamma_derivative-5}
\end{eqnarray}
By substituting Eqs.~(\ref{Appendix_Gamma_derivative-1})--(\ref{Appendix_Gamma_derivative-5}) into Eq.~(\ref{Appendix-RG_Delta2_dimensionfull-1}), one can obtain the flow equation (\ref{RG_Delta2_dimensionfull-1}).
For the terms $(\mathrm{B}-1)$, $(\mathrm{B}-2)$, $(\mathrm{B}-3)$, and $(\mathrm{C})$, the loop integral of the frequency $\omega$ in Eq.~(\ref{Appendix-RG_Delta2_dimensionfull-1}) is trivial because the functional derivatives contain the delta function $\delta(\omega)$.
In contrast, the term $(\mathrm{A})$ has a nontrivial frequency integral which is proportional to the temperature $T_k$ (see $(\mathrm{A})$ in Eq.~(\ref{RG_Delta2_dimensionfull-1})).

\end{document}